\documentclass[graybox, envcountchap]{svmult}

\usepackage{mathptmx}        
\usepackage{amsmath}
\usepackage{amssymb}
\usepackage{color}
\usepackage{helvet}          
\usepackage{courier}         
\usepackage{dirtree}

\usepackage{makeidx}        
\usepackage{graphicx}        
\usepackage{subfig}

\usepackage{multicol}        
\usepackage[bottom]{footmisc}

\usepackage{hyperref}        
\hypersetup{colorlinks=true,urlcolor=blue,citecolor=blue}

\usepackage[misc]{ifsym}
\usepackage{xcolor}
\usepackage[numbers,sort&compress]{natbib}
\newcommand{\aap}{A\&A}

\newcommand{\apj}{ApJ}
\newcommand{\apjl}{ApJL}

\newcommand{\mnras}{MNRAS}
\newcommand{\nat}{Nature}
\newcommand{\pasj}{PASJ}
\newcommand{\pasp}{PASP}
\newcommand{\procspie}{Proceedings of the SPIE}
\newcommand{\ssr}{Space Science Reviews}

\makeindex             

\begin{document}


\title{XMM-Newton Reflection Grating Spectrometer}
\titlerunning{XMM-Newton/RGS}
\author{Junjie Mao, Frits Paerels, Matteo Guainazzi, Jelle S. Kaastra}
\institute{Junjie Mao (\Letter) \at Department of Astronomy, Tsinghua University, Haidian DS 100084, Beijing, People’s Republic of China. \email{jmao@tsinghua.edu.cn}
\and Frits Paerels \at Columbia Astrophysics Laboratory, 550 West 120th Street, New York, NY 10027, USA. \email{frits@astro.columbia.edu}
\and Matteo Guainazzi \at ESA European Space Research and Technology Centre (ESTEC), Keplerlaan 1, 2201 AZ, Noordwijk, the Netherlands. \email{Matteo.Guainazzi@sciops.esa.int}
\and Jelle S. Kaastra \at SRON Netherlands Institute for Space Research, Niels Bohrweg 4, 2333 CA Leiden, the Netherlands.
\email{J.Kaastra@sron.nl}
}
%
%
\maketitle

\abstract{The past two decades have witnessed the rapid growth of our knowledge of the X-ray Universe thanks to flagship X-ray space observatories like XMM-Newton and Chandra. A significant portion of discoveries would have been impossible without the X-ray diffractive grating spectrometers aboard these two space observatories. We briefly overview the physical principles of diffractive grating spectrometers as the background to the beginning of a new era with the next-generation (diffractive and non-diffractive) high-resolution X-ray spectrometers. This chapter focuses on the Reflection Grating Spectrometer aboard XMM-Newton, which provides high-quality high-resolution spectra in the soft X-ray band. Its performance and excellent calibration quality have allowed breakthrough advancements in a wide range of astrophysical topics. For the benefit of new learners, we illustrate how to reduce RGS imaging, timing, and spectral data. }



\section{Introduction}

The past two decades have witnessed the rapid growth of our knowledge of the X-ray Universe thanks to flagship X-ray space observatories like XMM-Newton and Chandra \citep{Santos-Lleo2009,Wilkes2022}. A significant portion of discoveries would have been impossible without the X-ray diffractive grating spectrometers aboard these two space observatories \citep{Cottam2002,Drake2005,Miller2006,Neilsen2009,Kaastra2014, Miller2015,Pinto2016a,Nicastro2018,Argiroffi2019,Miceli2019,Shi2021}. We briefly overview the physical principles of diffractive grating spectrometers as the background to the beginning of a new era with the next-generation (diffractive and non-diffractive) high-resolution X-ray spectrometers (Section~\ref{sct:xray_spec_inst}). Then we focus on the Reflection Grating Spectrometer aboard XMM-Newton, which provides high-quality high-resolution spectra in the soft X-ray band (Section~\ref{sct:xmm_rgs}). Its performance and excellent calibration quality \citep{deVries2015} have allowed breakthrough advancements in a wide range of astrophysical topics \citep{Kahn2001,Rasmussen2001,Cottam2002, Kinkhabwala2002,Peterson2003,Pounds2003,Branduardi-Raymont2007,Miller2013,Miller2015,Pinto2016a,Ogorzalek2017,Nicastro2018,Mao2021}. For the benefit of new learners, we illustrate how to reduce RGS imaging, timing, and spectral data (Section~\ref{sct:rgs_data}).

\section{Diffractive and non-differactive X-ray spectrometers}
\label{sct:xray_spec_inst}
Generally speaking, there are two types of X-ray spectrometers: diffractive and non-diffractive. We briefly explain the physical principle of non-diffractive spectrometers in Section~\ref{sct:non_diff_inst}, including both traditional Charge-Coupled Devices (CCD) and revolutionary micro-calorimeters. The latter provides spectra (at very high resolution in the case of micro-calorimeters) based on different physical principles in a distinct principle when compared to diffractive spectrometers. In Sections~\ref{sct:diff_inst0}, we briefly describe diffractive crystal and grating spectrometers. The latter is widely used since it can obtain spectra over a wide energy interval simultaneously. Then we illustrate the impact of imperfections in the optical arrangement of diffraction gratings (Section~\ref{sct:diff_not_perfect}). Future designs for diffractive grating spectrometers are presented in Section \ref{sct:diff_inst1}. Both diffractive and non-diffractive designs have advantages and trade-offs (Section~\ref{sct:diff_or_not}).

\subsection{Non-diffractive X-ray spectrometers}
\label{sct:non_diff_inst}
Widely used Silicon-based Charge-Coupled Devices (CCDs) are non-diffractive spectrometers, such as the European Photon imaging camera (EPIC) \cite{Struder2001,Turner2001} aboard XMM-Newton \cite{Jansen2001}, the Advanced CCD imaging camera (ACIS) \cite{Garmire2003} aboard Chandra \cite{Weisskopf2002}, and the X-ray Imaging Spectrometer (XIS) aboard Suzaku \cite{Koyama2007}. The energy of the incident photon is absorbed by silicon to generate a primary (photon-induced) electron, followed by collisional processes to produce more electrons \citep{Paerels1999}. The average number of photon-generated electrons is \citep{Paerels1999}
\begin{equation}
\label{eq:ne_non_diff}
    N_e=E_\gamma/w,~
\end{equation}
where $E_\gamma$ is the energy of the incident photon and $w$ is the average energy required to give rise to one electron-hole pair. The $w$ parameter varies for different detector materials and operating temperatures. For Si-based detectors operating at $-100~^{\circ}$C, $w=3.68$~eV \citep{Fraser1994}. Accordingly, for an incident photon with $E_\gamma=6$~keV, the average number of photon-generated electrons is $\sim1630$. The variance of the number of photon-generated electrons is not given by the normal Poisson statistics due to the correlation between the energy-recovery processes \citep{Fraser1994}. The variance of the number of photon-generated electrons is \citep{Paerels1999}
\begin{equation}
    \Delta N_e=\sqrt{F~N_e}=\sqrt{F~E_\gamma/w}
\end{equation}
where $F(<1)$ is the fano factor \citep{Fraser1994}. For Si-based detectors operating at $-100~^{\circ}$C, the fano factor is $0.11$ \citep{Fraser1994}. The maximum (``Fano limited") energy resolution is then \citep{Paerels1999}
\begin{equation}
\label{eq:delta_eng_fano}
    \Delta E = 2.35 w \sqrt{F~E_\gamma/w}.
\end{equation}
For Si-based detectors operating at $-100~^{\circ}$C, $\Delta E\sim47(E_\gamma/{\rm keV})^{0.5}~{\rm eV}$. According to Eq.~\ref{eq:delta_eng_fano}, the higher the incident photon energy ($E_\gamma$), the larger the energy resolution ($\Delta E$). The energy resolution is a slow function of $E_{\rm \gamma}$ though. In practice, the total energy resolution should take system noise ($\sigma$) into account \citep{Holland2010}, 
\begin{equation}
\label{eq:delta_eng_total}
    \Delta E = 2.35 w \sqrt{F~E_\gamma/w + \sigma^2}.
\end{equation}
When the instrument is operated at $\sim-100~^{\circ}$C, system noise contributed by leakage current is negligible. At high readout frequencies, in the absence of other degrading factors, the on-chip amplifier white noise ($\sigma_{\rm w}$) dominates the system noise \citep{Holland2010}. In this case, $\sigma_{\rm w}\propto \sqrt{f_{\rm ro}}$, where $f_{\rm ro}$ is the readout frequency \citep{Holland2010}. 

The energy resolution of Chandra/ACIS spectroscopic array is $\sim95$~eV at 1.5~keV and $150$~eV at 5.9~keV. But the energy resolution of a non-diffractive spectrometer is not necessarily low. Micro-calorimeter, another type of non-diffractive spectrometer, can achieve an energy resolution of $\Delta E\lesssim10$~eV, which is an order of magnitude better than ACIS and EPIC. 

A micro-calorimeter unit consists of three basic components: an X-ray absorber, a sensitive temperature sensor, and a weak thermal link \citep{McCammon2002}. After absorbing the incident X-ray photon, the detector temperature will experience a sharp rise followed by a slow decay. The temperature sensor will convert the temperature pulse to an electronic pulse (to be read out). The weak thermal link will remove excess heat in preparation for absorbing the next incident photon. The energy resolution of the device is \citep{Cui2020b}
\begin{equation}
    \Delta E = \eta \sqrt{\frac{kT^2C}{\alpha}},~
\end{equation}
where $\eta$ is a dimensionless parameter of the
order of unity, $k$ the Boltzmann constant, $T$ the detector temperature ($\sim50$~mK, slightly above absolute zero), $C$ the heat capacity, and $\alpha$ the temperature sensitivity. High energy resolution can be achieved by reducing the heat capacity and increasing the temperature sensitivity. We refer readers to Chapters \textcolor{red}{X} (Micro-calorimeters with transition-edge sensors) and \textcolor{red}{Y} (Hitomi/XRISM micro-calorimeter) of this book for more technical details. 

The first astrophysical application of micro-calorimeter was conducted by \citet{McCammon2002} on a sounding rocket launched in 1999. Its energy resolution ranges from $5-12$~eV over the $0.06-1$~keV band \citep{McCammon2002}. This micro-calorimeter was Silicon-based. The same design was adopted by Hitomi/SXS (Soft X-ray Spectrometer) \citep{Hitomi2016} and XRISM/Resolve \citep{XRISM2020}. Micro-calorimeter based on transition-edge sensors (TES) has an even better performance. For instance, Athena X-ray observatory is an approved ESA large class mission \citep{Nandra2013}. Its X-ray Integral Field Unit (X-IFU) \citep{Barret2018} aims to achieve an energy resolution of $2.5$~eV over the $0.2-12$~keV energy band. The Hot Universe Baryon Surveyor (HUBS) \citep{Cui2020a,Cui2020b}, a mission proposed to the Chinese National Space Agency (CNSA), aims to achieve an energy resolution of 0.6~eV for its central $12\times12~{\rm pixel^2}$ over the energy band of $0.1-2$~keV. 

The energy resolution ($\Delta E$) of CCDs and micro-calorimeters is constant or only moderately dependent on energy. Accordingly, their spectra are often plotted in the energy space. 

\subsection{Past and current diffractive X-ray spectrometers}
\label{sct:diff_inst0}
The aforementioned non-diffractive spectrometers convert the incident photon energy to countable objects (e.g., electrons). Diffractive spectrometers construct interference of incident photons along different light paths based on their photon energy. This is realized by placing a diffraction element at the exit aperture of the focusing optic. Dispersed X-ray photons will then arrive at a focal plane imaging detector (usually a non-diffractive imaging spectrometer). 

The first astrophysical application of a diffractive X-ray spectrometer is actually a crystal spectrometer, i.e., the Focal Plane Crystal Spectrometer (FPCS) on the Einstein Observatory \citep{Canizares1979}. FPCS has a resolving power $R=E/\Delta E=50-500$ over the energy band of $0.2-3$~keV \citep{Canizares1979}. Limited by the Bragg condition, crystal spectrometers cannot simultaneously provide diffract radiation for a wide range of wavelengths. To overcome this drawback, the device needs to scan through a range of Bragg angles (by rocking the FPCS crystal) \citep{Canizares1979}. Crystal spectrometers were also used on the Solar Maximum Mission \cite{Acton1980}, HINOTORI \cite{Kondo1982}, P78-1 \cite{Doschek1983}, SOLAR-A \cite{Culhane1991}, and YOHKOH \cite{Lang1992}. In particular, the Bragg Crystal Spectrometer aboard SOLAR-A achieved a resolving power $R=\lambda/\Delta \lambda=3000-6000$ for some narrow wavelength ranges. 

Unlike crystal spectrometers, grating spectrometers can obtain spectra over a wide energy interval simultaneously. The Reflection Grating Spectrometer (RGS) \cite{denHerder2001} aboard XMM-Newton, High- and Low-Energy Transmission Grating Spectrometers (HETGS \cite{Canizares2005} and LETGS \cite{Brinkman2000}) aboard Chandra are the current main working horses to obtain high-resolution X-ray spectra. As can be told from the names of these instruments, classical grating spectrometers can be further divided into two types: transmission and reflection gratings. 

\begin{figure}
\centering
\includegraphics[width=\hsize, trim={0.5cm 2.0cm 1.0cm 3.5cm}, clip]{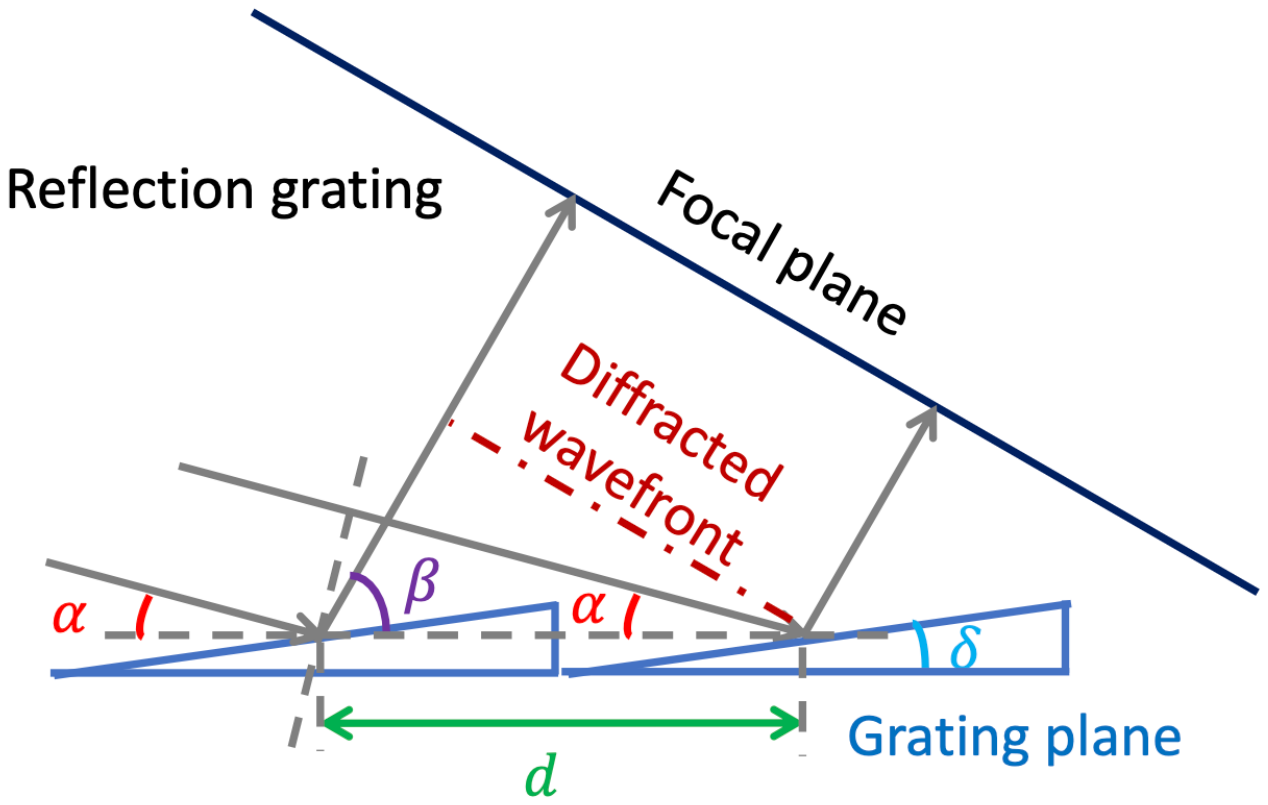}
\caption{Cartoon of the dispersion geometry for the reflection gratings. X-rays enter from the left with an incidence angle $\alpha$ with respect to the grating plane. The triangular grating grooves are titled by $\delta$ with respect to the grating plane.}
\label{fig:rgs_diff}
\end{figure}

Prior to XMM-Newton, reflection gratings were applied to the Extreme UltraViolet Explorer (EUVE) mission \citep{Bowyer1991}. The three reflection gratings cover three UV wavelength ranges: $70-190$~\AA, $140-380$~\AA, and $280-760$~\AA, respectively. In addition, reflection gratings were used on the recent sounding rocket instrument -- Marshall Grazing Incidence X-ray Spectrometer \cite{Champey2022}. Prior to Chandra, transmission gratings were applied to Einstein  \citep{Seward1982} and EXOSAT \citep{Brinkman1980}. 

For reflection gratings, the dispersion relation is \citep{denHerder2001}
\begin{equation}
\label{eq:disp_refl}
    m \lambda = d (\cos \beta - \cos \alpha),~
\end{equation}
where $m=0,~\pm1,~\pm2,~...$ is the spectral order, $\lambda$ the wavelength of the incident photon, $d$ the grating period, $\beta$ the dispersion angle, $\alpha$ the incident angle on the grating plane. Note that triangular grooves on the grating plane are tilted by $\delta$ (Fig.~\ref{fig:rgs_diff}). For the zeroth order, we have $\alpha_{m=0}=\beta_{m=0}$, and the grating acts as a mirror. For negative orders, also known as inside orders, we have $\alpha<\beta<\beta_{m=0}$. Positive orders, also known as outside orders, might not always exist, because $\cos\beta$ might be greater than unity for some combinations of $\alpha$, $d$, and $\lambda$. In the absence of other degrading factors, the wavelength resolution of reflection gratings is \citep{Paerels2010} 
\begin{equation}
\label{eq:delta_lamb_refl}
    \Delta \lambda = \frac{d}{m} \sin \alpha \Delta \alpha,~
\end{equation}
where $\Delta \alpha$ is the angular resolution of the telescope. 

\begin{figure}
\centering
\includegraphics[width=.8\hsize, trim={0.5cm 11.0cm 1.0cm 7cm}, clip]{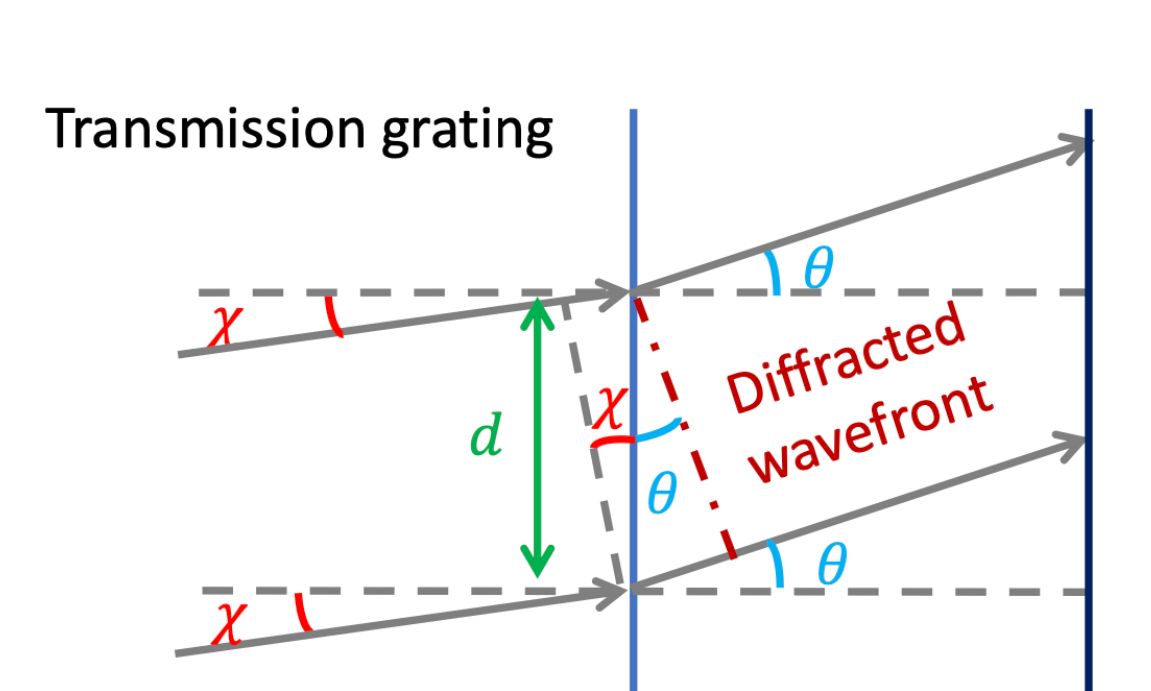}
\caption{Cartoon of the dispersion geometry for the transmission gratings. X-rays enter from the left with an incidence angle $\chi$ with respect to the normalization direction of the grating plane. }
\label{fig:tgs_diff}
\end{figure}

For transmission gratings, the dispersion relation is \citep{Paerels2010}
\begin{equation}
\label{eq:disp_tran}
    m \lambda~ = d (\sin \theta - \sin \chi)~,
\end{equation}
where $m=0,~\pm1,~\pm2,~...$ is the spectral order, $\lambda$ the wavelength of the incident photon, $d$ the grating period, $\theta$ the dispersion angle, $\chi$ the incident angle on the grating (Fig.~\ref{fig:tgs_diff}). Since the incidence and dispersion angles are often small, the wavelength resolution is \citep{Paerels2010} 
\begin{equation}
\label{eq:delta_lamb_tran}
    \Delta \lambda = \frac{d}{m}\Delta (\theta - \chi)
\end{equation}

Both reflection and transmission gratings are constant $\Delta \lambda$ devices (Eq.~\ref{eq:delta_lamb_refl} and \ref{eq:delta_lamb_tran}). Accordingly, grating spectra are often plotted in the wavelength space. 

Note that for any slitless spectrometer, the incident and dispersion angles will vary across a finite-size grating placed in a focused beam. In practice, the Rowland circle was used to refocus the radiation by curving the gratings \citep{Paerels1999,Paerels2010}. Misalignment of the gratings or deviations from the flatness of the grating plane can also play a significant role in the performance of the gratings (\citep{Paerels1999}, see also Section~\ref{sct:diff_not_perfect}). 

\subsection{Scattering by diffraction gratings}
\label{sct:diff_not_perfect}
Scattering by diffraction gratings is a natural consequence of imperfections in the optical arrangement, whether by roughness on the surface of a grating, or by any random variability of any of the grating properties (such as groove spacing). If this variability occurs on spatial wavelength scales that are short compared to the grating period, the main effect will be an apparent loss of light from the brightest diffraction peaks, without a detectable effect on the spatial response function (i.e., the line spread function, LSF). If, on the other hand, there is variability on spatial wavelength scales comparable to the grating period, the LSF will be affected and appear to have significant scattering `wings'. In the following, we will develop a very simple description of scattering that will allow for an intuitive understanding and provide useful first-order quantitative scalings. We will describe the scattering properties of the XMM-Newton/RGS, but start with a simple example based on transmission gratings to introduce the analytical problem.

We will assume, as usual, that the grating period, $d$, is much larger than the radiation wavelength $\lambda$. We will also ignore the detailed interaction between the grating material and the radiation, meaning, we will ignore the vector character of the electromagnetic field and the boundary conditions on the solutions to Maxwell's equations. In the approximation that the diffracted radiation is detected far from the diffracting element, we will just apply Huygens' Principle in a slightly extended formulation: we will sum the spherical waves emitted along the wavefronts at the diffraction grating, possibly with an amplitude and phase that have been modulated by propagation through the material.

Let us assume a very simple model for a grating: a set of points at locations ${\bf r}_i$, periodically spaced, each of which radiates spherical waves in response to being illuminated by a plane wave
\begin{equation}
A_{\rm in} = \exp i {\bf k}_{\rm in} \cdot {\bf r}
\end{equation}
with ${\bf k}_{\rm in}$ the wave vector of the incident waves.
Far from the grating, the diffracted wave propagating in direction ${\bf k}_{\rm out}$ has complex amplitude (ignoring an uninteresting constant phase factor)
\begin{equation}
A = \sum_i \exp i ({\bf k}_{\rm in} -{\bf k}_{\rm out})\cdot {\bf r}_i
\end{equation}
and the intensity of the diffracted light will be given by $I = \left| A \right|^2$. In the following we will set ${\bf k}_{\rm in} -{\bf k}_{\rm out} = {\bf q}$. The mathematically perfect grating therefore produces a diffraction pattern
\begin{equation}
I_0 = AA^* = \left | \sum_i \exp i {\bf q}\cdot {\bf r}_i  \right|^2
\end{equation}

Now imagine that the location of each radiating point is slightly displaced by a random displacement ${\bf r}_i \rightarrow {\bf r}_i + \delta{\bf r}_i$, with $\delta r_i / r_i \ll 1$ and $\left< \delta {\bf r}_i  \right> = 0$. The $\delta{\bf r}_i$ are uncorrelated, so that when averaged over the entire grating $\left<  \delta{\bf r}_i \cdot \delta{\bf r}_j \right> = 0$. The variance of the displacements is $\left<  \delta{\bf r}_i \cdot \delta{\bf r}_i \right> = \sigma^2$. Calculating the intensity of the diffracted field, and expanding to the first nonzero new term, we get
\begin{eqnarray}
\label{eq:i_diff_field}
I & = & \left< AA^* \right>  \cr
& = & 
\left<   \sum_i \exp i {\bf q}\cdot ({\bf r}_i +  \delta{\bf r}_i) \cdot 
\sum_j \exp -i {\bf q}\cdot ({\bf r}_j +  \delta{\bf r}_j)  \right> \cr
& = & \left<  \sum_i \sum_j \exp i {\bf q} \cdot \left(  {\bf r}_i - {\bf r}_j \right)  
\exp i {\bf q} \cdot \left(  \delta{\bf r}_i - \delta{\bf r}_j \right)  \right>  \cr
& & \cr
& \approx &  \sum_i \sum_j \exp i {\bf q}\cdot ({\bf r}_i - {\bf r}_j)  \cdot \cr
& &\ \  \left<   1 +   i {\bf q}\cdot ({\bf r}_i - {\bf r}_j) - {{1}\over{2}}q^2 \left(   \delta{\bf r}_i \cdot \delta{\bf r}_i + \delta{\bf r}_j \cdot \delta{\bf r}_j 
-2 \delta{\bf r}_i \cdot \delta{\bf r}_j 
        \right) + ...  \right> \cr
& & \cr
        & = & \sum_i \sum_j \exp i {\bf q}\cdot ({\bf r}_i - {\bf r}_j)  \cdot \left( 1 - q^2\sigma^2 + ...  \right) \cr
& & \cr
        & = & I_0  \left( 1 - q^2\sigma^2 + ...  \right) 
\end{eqnarray}

As long as $q^2\sigma^2 \ll 1$, the only effect of the perturbations is to remove a small amount of light, $I/I_0 = q^2\sigma^2$ from the sharp diffraction pattern, and distribute it widely in between the diffraction peaks (moving the lattice points ${\bf r}_i$ around conserves the number of photons, so whatever is missing from the diffraction peaks must have ended up in between the peaks).

A straightforward application of this idea to small perturbations to the period of a transmission grating is given by \cite{Davis1997} and \cite{Paerels1997}. Monochromatic line radiation 
dispersed by the High Energy Grating that is part of Chandra/HETGS onto a CCD camera showed that there appeared to be a faint continuum between the sharp diffraction orders. Inspection of the CCD spectrum of these photons showed that in fact they were all of the same energy, indicating that they had ended up dispersed far away from the diffraction maxima. The same thing was seen in dispersed light from Chandra/LETGS.

\begin{figure}
\centering
\includegraphics[width=1.0\hsize, trim={0.5cm 0.0cm 0.5cm 0.5cm}, clip]{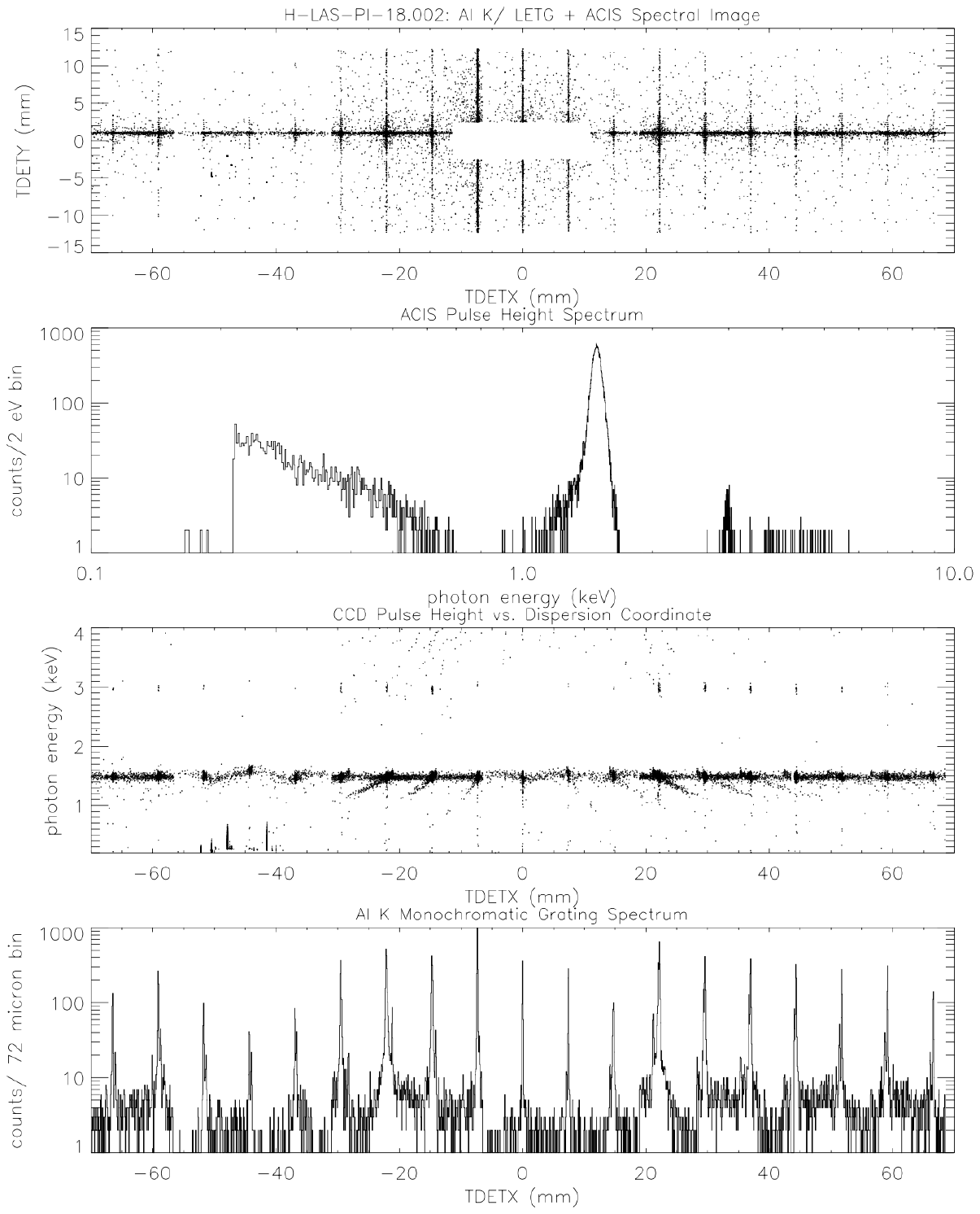}
\caption{Al K$\alpha$ $\lambda 8.34$~\AA\ radiation diffracted by the {\it Chandra} Low Energy Transmission Grating in a ground calibration exposure. Data were recorded with a CCD camera. The top panel shows the spatial diffraction pattern. The second panel shows the CCD spectrum; the third panel shows photon energy versus dispersion coordinate (TDETX) for each photon. The bottom plot shows the diffraction pattern, filtered to exclude continuum photons, integrated over cross-dispersion coordinate TDETY. Data courtesy of Dan Dewey (MIT).
}
\label{fig:letg_scatter}
\end{figure}

A calculation for the diffraction pattern in the presence of small perturbations to the grating period shows that the fraction of scattered light is 
\begin{equation}
f = q^2\sigma^2 = k^2\sigma^2 \sin^2\theta = 4\pi m^2 \left({{\sigma}\over{d}}\right)^2
\end{equation}
where we assume the radiation is incident perpendicularly to the grating, the dispersion angle $\theta$ is given by $\sin\theta = m\lambda/d$, and $m$ is the diffraction order, $d$ the average grating period. There is no effect in the zeroth order, and the effect grows quadratically with diffraction orders. Fig.~\ref{fig:letg_scatter} shows the diffraction pattern of the LETG of Al K$\alpha$ radiation (8.34 \AA), out to $m=9$. The bottom graph shows the spectrum in a narrow band of energies centered on 8.34 \AA\ (so the faint continuum visible as curved bands in the third panel from the top has been filtered out). The photons in between the diffraction peaks are now almost all Al K$\alpha$, and this `continuum' clearly rises sharply with diffraction order. This graph incidentally also shows that the Al K$\alpha$ source is not strictly monochromatic; in the third order especially a second emission line is clearly visible on the high energy side of Al K$\alpha$. Most likely this is Al K$\alpha$ excited in aluminum oxide present on the electron impact source!

In order to understand the spatial distribution of the scattered light, imagine that the displacements $\delta{\bf r}_i$ are expanded in a Fourier series or integral. Assuming at first just a single sinusoidal perturbation, $\delta{\bf r}_i = {\bf w}\sin ({\bf g}\cdot{\bf r}_i)$, with ${\bf g}$ the wavevector of the sinusoidal perturbation, of period $l = 2\pi/|{\bf g}|$. Introducing this into Eq.~\ref{eq:i_diff_field} and carrying through the calculation, assuming that the amplitude $w$ of the perturbation is small (so the Bessel functions of argument ${\bf q}\cdot{\bf w}$ that will appear can be expanded to first order), it is straightforward to show that the resulting diffraction pattern, in addition to peaking at $\sin\theta = m\lambda/d$, also peaks at $\sin\theta = m\lambda/d \pm \lambda/l$ with an amplitude proportional to $|{\bf q}\cdot{\bf w}|^2$. These are the first-order terms; we ignore the higher-order terms. The wave-like perturbation will cause diffracted light to appear at an angle $\Delta\theta \approx \lambda/l$ away from the main diffraction order. Now imagine that the perturbation is made up of an ensemble of sine waves of wave vectors $g$, with a power spectrum $W(g)$. In the first order, each sine wave contributes diffracted light at an angle $\Delta\theta = \lambda g/2\pi$ with respect to a given diffraction order, with an amplitude determined by $W(g)$. The resulting angular distribution of scattered light therefore directly maps out the power spectral distribution of the perturbations. 

This can be applied to the reflection gratings on the RGS. An abbreviated calculation is given by \cite{Paerels1994, Kahn1996, Paerels2001}. We expect to see the effect of variations in the grating properties (period, groove profile) as well as scattering by surface roughness. We based our analysis of scattering in RGS on the assumption that scattering by surface roughness is probably dominant. Scattering by variations in groove profile and period are not likely to be dominated by coherent perturbations on the scale of around a few dozen grooves or less (in which case they will cause a distribution of scattered light comparable in width to the LSF), but instead light scattered by groove profile variations probably simply merges with the surface roughness scattering, at wider angles. 

We pursued a scalar calculation for the grating diffraction pattern analogous to the one described above for the case of a transmission grating. The angle of incidence on the grating is $\alpha$, the dispersion angle is $\beta$, the grating period is $d$, and the dispersion relation for diffraction order $m$ is given by
\begin{equation}
\cos\beta_m - \cos\alpha = {{m\lambda}\over{d}}
\end{equation}
In the case of the gratings and the geometry on RGS, only orders $m \leq 0$ exist. The scalar perturbation theory now gives for the intensity of light dispersed at angle $\beta$ out of diffraction order $m$:
\begin{equation}
{{1}\over{I_m}}{{dI_m}\over{d\beta}} = {{(\sin\alpha + \sin\beta)^4}\over{(\sin\alpha + \sin\beta_m)^2}} \sin\beta_m\ k^3 W(p)
\end{equation}
Here, $I_m$ is the intensity in diffraction order $m$ of light of wavelength $\lambda = 2\pi/k$, and $p = k(\cos\alpha - \cos\beta^{\pm}) \approx
\pm k\sin\beta_m\Delta\beta$, and $\Delta\beta$ is the angle $\beta-\beta_m$. The total fraction of light $f = I_{\rm scattered}/I_m$ scattered off order $m$ is 
\begin{equation}
f = k^2\sigma^2 (\sin\alpha+\sin\beta_m)^2
\end{equation}
with $\sigma^2$ the variance of the surface roughness. The power spectrum normalization is $\int W(p) dp = \sigma^2$.

\begin{figure}[!t]
\centering
\includegraphics[width=1.1\hsize, trim={3.0cm 6.0cm 0.5cm 0.5cm}, clip]{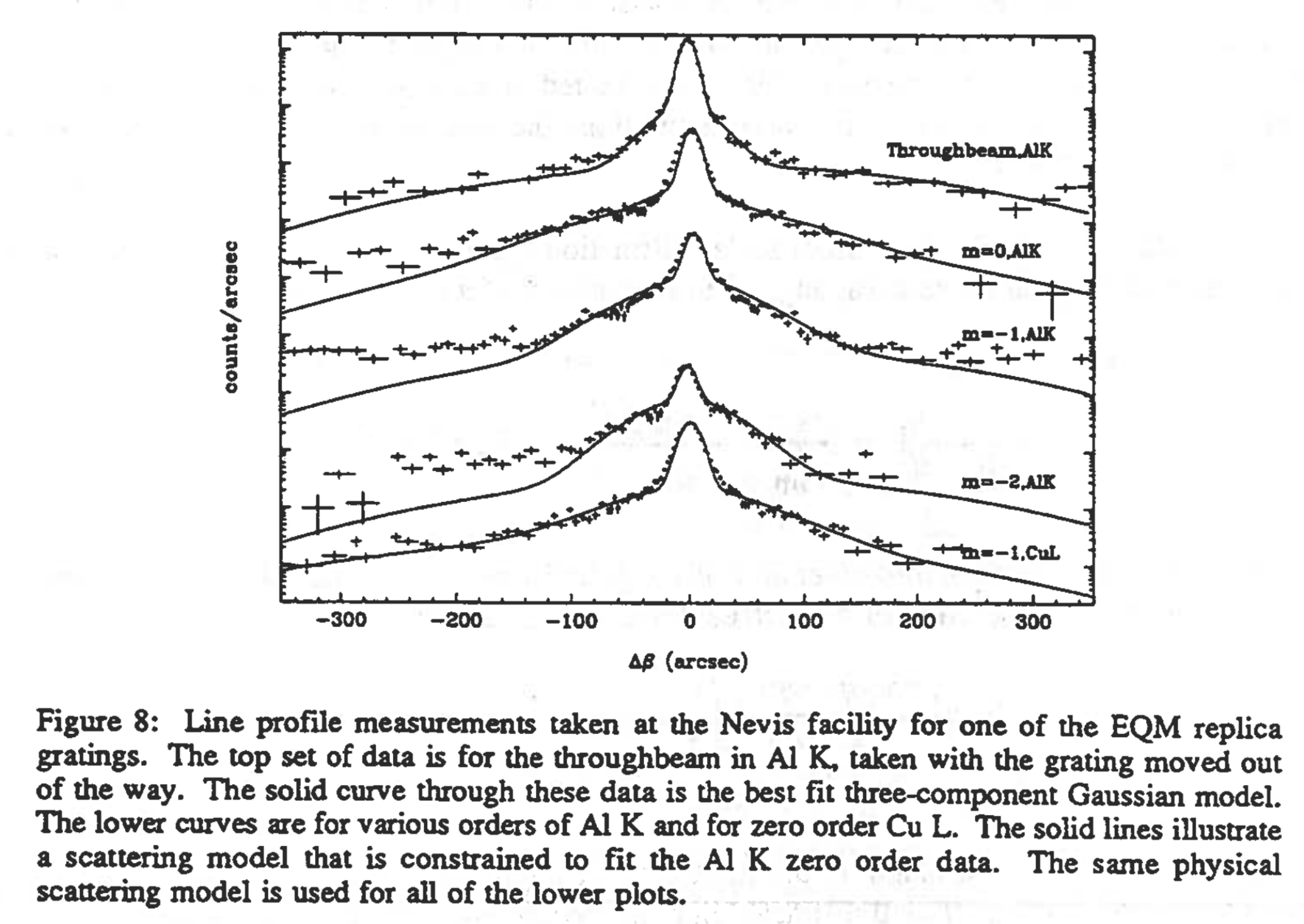}
\caption{Line profile measurements taken at the Nevis facility for one of the Engineering Qualification Model (EQM) replica gratings. The top set of data is for the throughbeam in Al K, taken with the grating moved out of the way. The solid curve through the data is the best-fit three-component Gaussian model. The lower curves are for various orders of Al K and for zero order Cu L. The solid lines illustrate a scattering model is used for all the lower plots. Figure reproduced with permission from \cite{Kahn1996}, SPIE. 
}
\label{fig:gsca_RGS}
\end{figure}

\begin{figure}[!h]
\centering
\includegraphics[width=0.9\hsize, trim={0.5cm 0.0cm 0.5cm 0.cm}, clip]{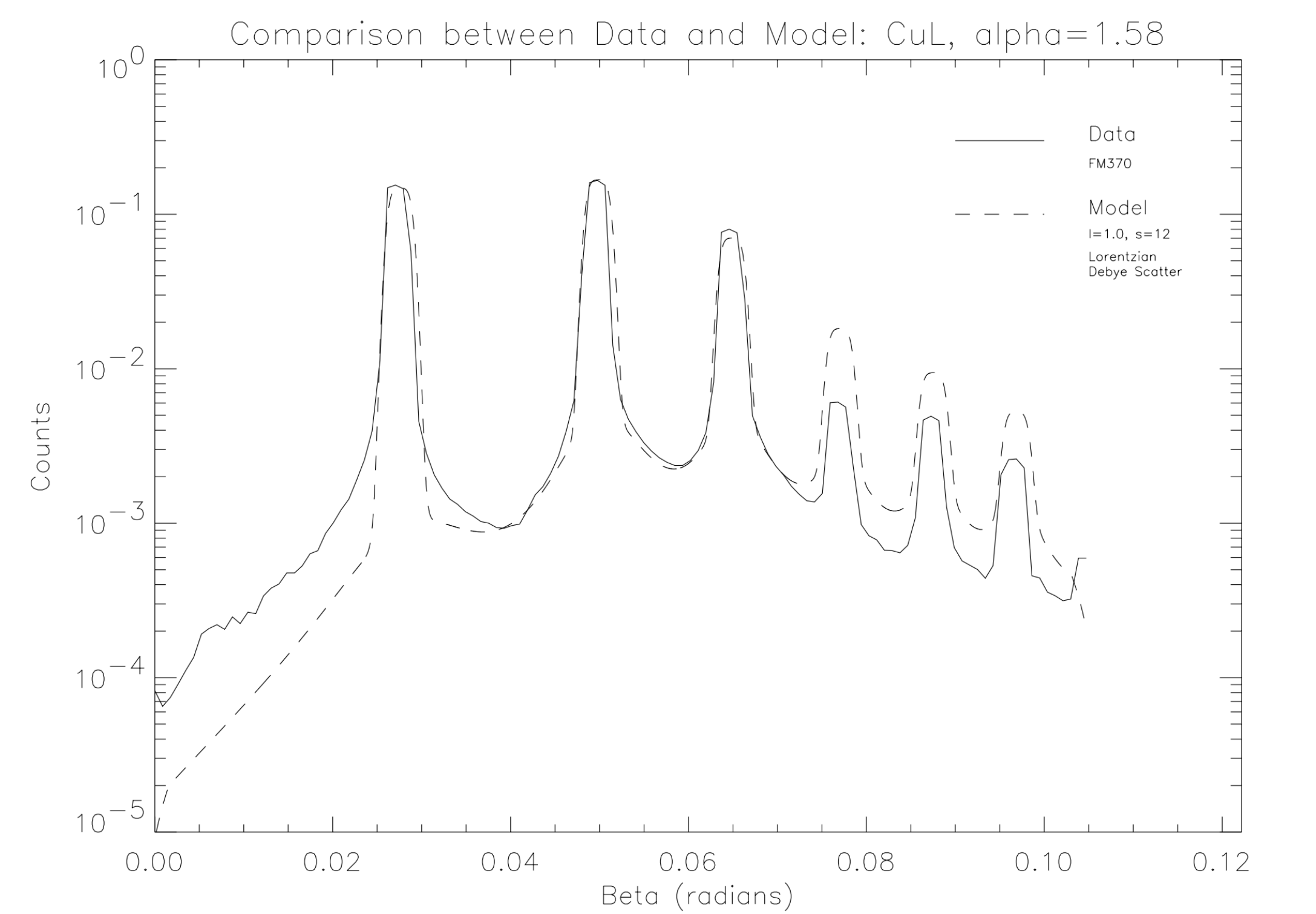}
\caption{The solid line shows data taken in Cu L radiation (13.33 \AA) at Columbia University's X-ray calibration facility at Nevis Laboratories for grating FM370.
The dashed line is a model based on large-angle scattering for 13.33 \AA\ photons, assuming a Lorentzian distribution of correlation length $l=1\mu$m and surface roughness $\sigma = 12$\AA. The peak at $\beta = 0.0275$ radians $ = 1.58$ deg is the zero order. Its profile is used to model through the beam for the higher orders. 
}
\label{fig:gsca_large_angle}
\end{figure}

\begin{figure}[!h]
\centering
\includegraphics[width=0.47\hsize, trim={0.5cm 0.0cm 0.5cm 0.cm}, clip]{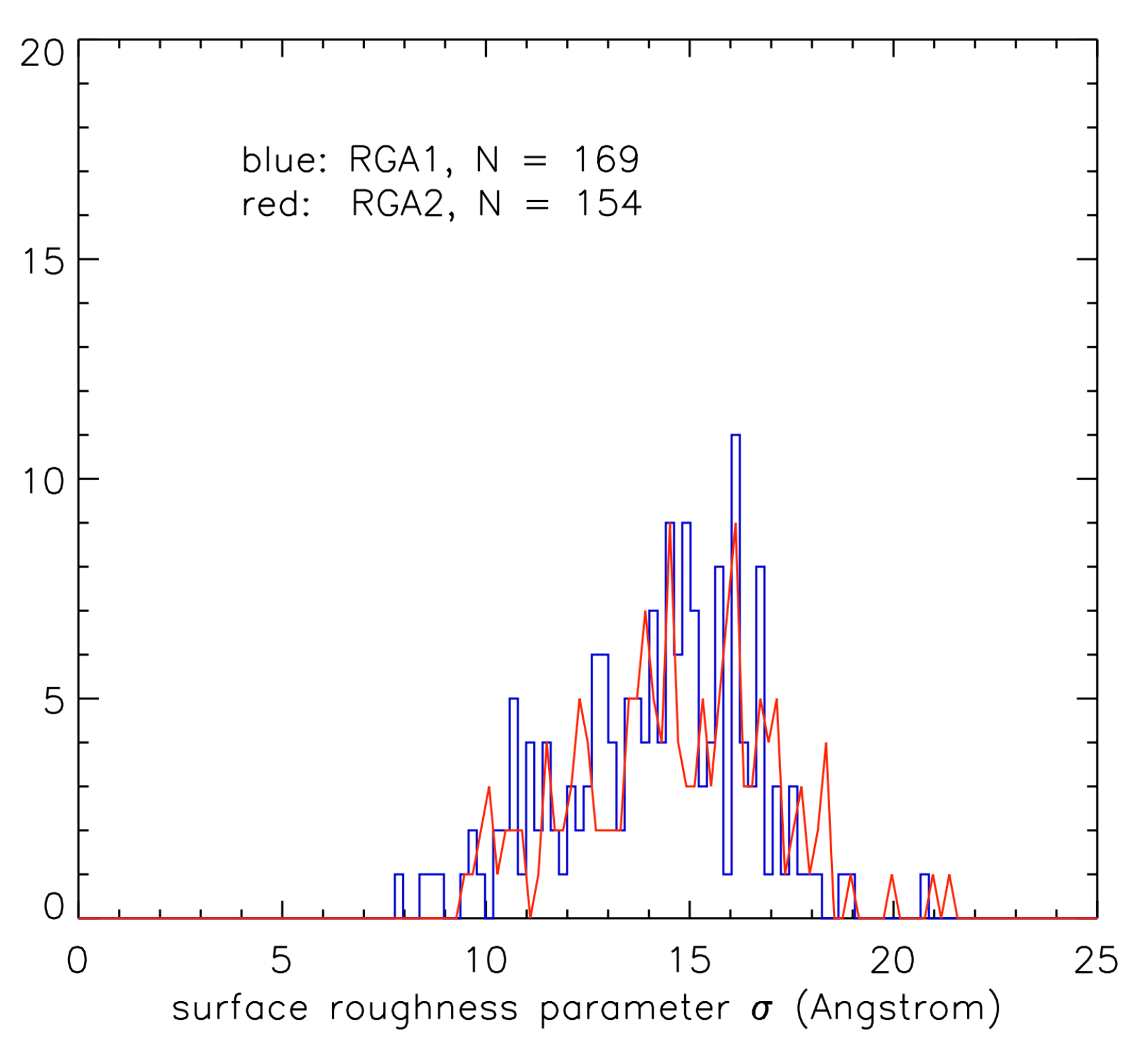}
\includegraphics[width=0.47\hsize, trim={0.5cm 0.0cm 0.5cm 0.cm}, clip]{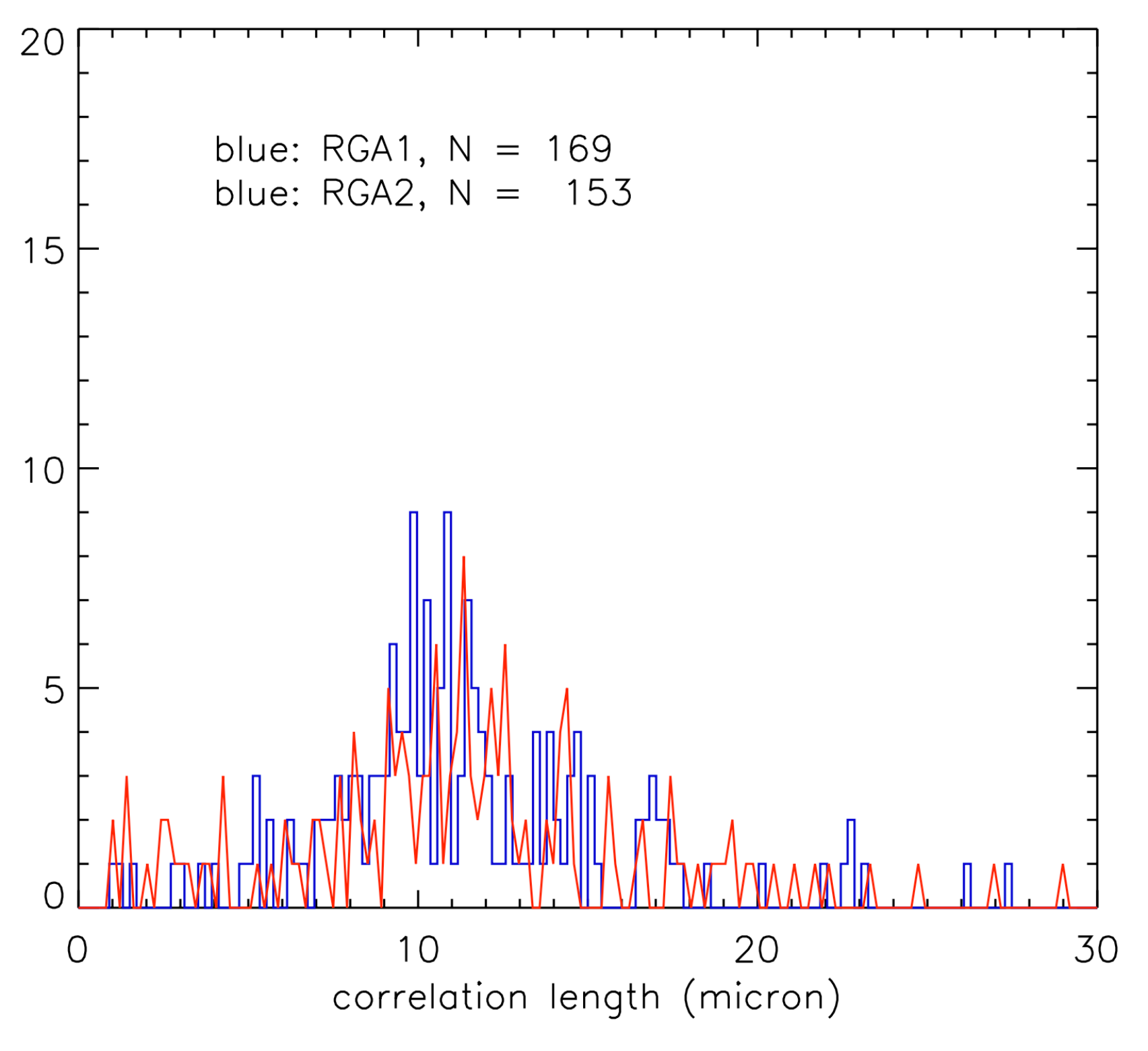}
\caption{Distribution of small-angle scattering surface roughness parameters (left) and correlation lengths (right) measured pre-flight for gratings that make up the two Reflection Grating Arrays (RGA).
}
\label{fig:srd_cld_rgs}
\end{figure}

In applying this to the LSF of the RGS, we found that there appeared to be two scattering distributions, of different spatial coherence lengths $l$, which we termed `large' and `small angle' scattering. The former effectively produces a correction on the diffraction efficiencies,
by scattering light far away from the diffraction peaks. The latter produces visible scattering `wings' to the LSF. Both should of course be taken into account in modeling the response of the spectrometer to a given incident spectrum (including the large-angle scattered light, which reduces apparent contrast). Fig.~\ref{fig:gsca_RGS} illustrates the small-angle scattering for one grating. Fig.~\ref{fig:gsca_large_angle} shows an `interorder scan' for one of the gratings, showing the effect of `large-angle scattering'. In the left panel of Fig.~\ref{fig:srd_cld_rgs} we show the distribution of surface roughnesses that gives rise to the small-angle scattering, for gratings that make up RGA1 and RGA2 (in RGS1 and RGS2, respectively). The corresponding distribution of correlation lengths is shown in the right panel of Fig.~\ref{fig:srd_cld_rgs}. For comparison, the average grating period of the gratings is $d = 1.54895~\mu{\rm m}$. 

These grating properties were incorporated into the flight instrument models, which were generated with ESA's {\tt SciSIM} package. Several documents describing the scattering calculations and the algorithms used to model the response exist (Cottam 1997a,b; available from frits@astro.columbia.edu on request).

\subsection{Future diffractive X-ray spectrometers}
\label{sct:diff_inst1}
The next generation of diffractive X-ray spectrometers have been proposed over the past few years: (1) Arcus is a soft X-ray grating spectrometer proposed to NASA as an Explorer Class mission \cite{Smith2019}; (2) HiReX is a medium-class mission proposed for ESA's Voyage 2050\footnote{https://www.cosmos.esa.int/web/voyage-2050}.
Table~\ref{tbl:cf_inst} compares the key parameters of grating spectrometers aboard XMM-Newton and Chandra, as well as the next generation of diffractive and non-diffractive spectrometers. Focusing on the grating spectrometers aboard XMM-Newton and Chandra, one yields a general impression that the latter has a better resolving power ($R$) but a smaller effective area ($A_{\rm eff}$). A desirable instrument should maximize the product of both $R$ and $A_{\rm eff}$. On one hand, we need a large resolving power to resolve closely spaced line features. On the other hand, we need a large effective area to collect more photons to increase the signal-to-noise ratio. For many weak absorption line studies \citep{Mao2017,Nicastro2018}, the figure of merit is $\sqrt{R~A_{\rm eff}}$ (see Technical note by Jelle Kaastra, Spectral diagnostics for IXO, https://space.mit.edu/home/nss/Jelle\_Kaastra\_ixo\_spextroscopy.pdf, \citep{Smith2019})\footnote{In the XRISM Quick Reference \cite{XRISM2022}, new figures of merit are defined. For strong lines with their equivalent width (EW) larger than the instrument resolution ($\Delta E$), figures of merit for line detection, bulk velocity, and line broadening are $\sqrt{A_{\rm eff}}$, $\sqrt{A_{\rm eff}~R^2}$, and $\sqrt{A_{\rm eff}~R^4}$, respectively. For weak lines (EW$\lesssim \Delta E$), we need to apply a factor of $\sqrt{1/\Delta E}$ to those of the strong lines.}. 

\begin{table}[h]
\centering
\small
\caption{Key parameters of the current and future high-resolution (diffractive or not) X-ray spectrometers. For grating spectrometers, only the first-order parameters are listed. For future missions, key parameters might be subjected to changes. The spectral resolution ($\Delta \lambda$ for diffractive devices and $\Delta E$ for non-diffractive devices), effective area $A_{\rm eff}$, resolving power $R=E/\Delta E=\lambda/\Delta \lambda$ and $\sqrt{A_{\rm eff}~R}$ are given at 0.5 keV.  }
\begin{tabular}{llllllll}
\noalign{\smallskip}
\hline\hline
\noalign{\smallskip}
Observatory & Spectrometer & Range & Resolution & $A_{\rm eff}$ & $R$ & $\sqrt{A_{\rm eff}~R}$  \\
\hline
\noalign{\smallskip}
XMM-Newton & RGS & $5-38$~\AA & $0.06$~\AA & 90 cm$^2$ & $400$ & 190 \\
\noalign{\smallskip}
Chandra & HEG (ACIS-S)  & $1.2-15$~\AA & $0.012$~\AA & $<1$ cm$^2$ & $2000$ & $<45$ \\
 & MEG (ACIS-S) & $2.5-31$~\AA & $0.023$~\AA & $5$ cm$^2$ & $1000$ & $70$ \\
 & LEG (HRC-S) & $1.2-175$~\AA & 0.05~\AA & 12 cm$^2$ & $500$ & $77$ \\
Arcus & -- -- & $10-50$~\AA & $0.0065$~\AA & 400 cm$^2$ & $3800$ & $1230$  \\ 
HiReX & -- -- & $8-124$~\AA & $0.0025$~\AA & 1500 cm$^2$ & $10^4$ & $3870$ \\
\noalign{\smallskip}
\hline
\noalign{\smallskip}
XRISM & Resolve & $0.3-12$~keV & $7$~eV & 125 cm$^2$ & $70$ & $94$ \\
HUBS & Central array & $0.1-2$~keV & $0.6$~eV & 400 cm$^2$ & $800$ & $570$ \\
    & Regular array & $0.1-2$~keV & $2$~eV & 300 cm$^2$ & $250$ & $270$ \\
Athena & X-IFU & $0.3-12$~keV & $2$~eV & 5900 cm$^2$ & $250$ & $1210$ \\
\noalign{\smallskip}
\hline
\end{tabular}
\label{tbl:cf_inst}
\end{table}

\begin{figure}[!b]
\centering
\includegraphics[width=.7\hsize, trim={0.5cm 0.0cm 0.5cm 0.5cm}, clip]{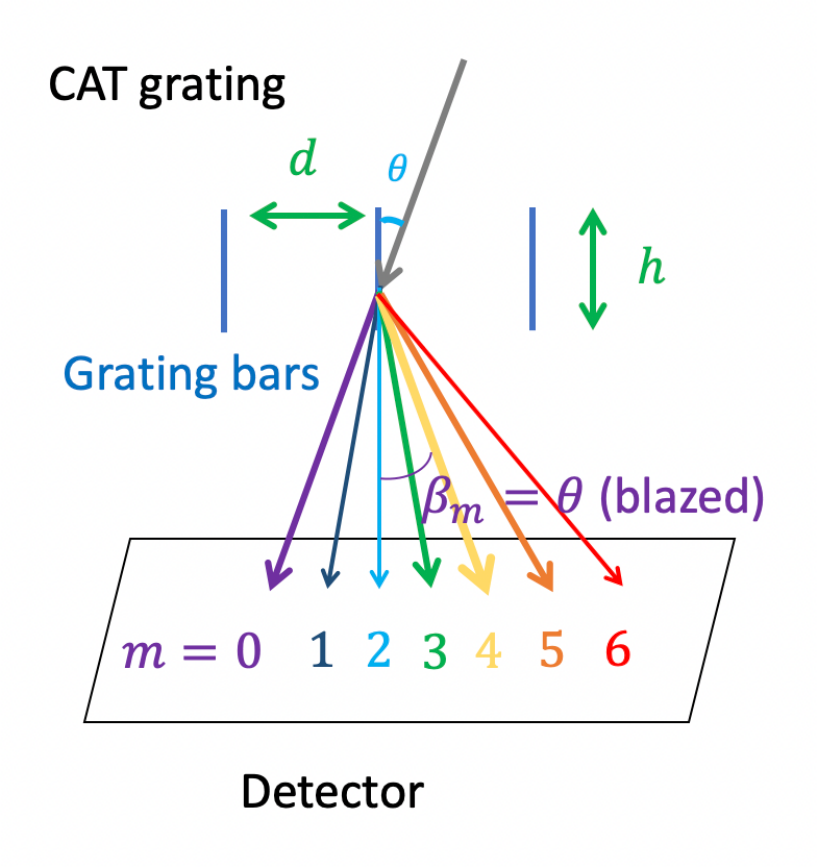}
\caption{Cartoon of the dispersion geometry of Critical Angle Transmission (CAT) grating (not to scale). X-rays enter from the top with an incidence angle $\theta$ with respect to the free-standing grating bar sidewalls. For $\theta$ smaller than a critical angle, efficient blazing enhances diffraction orders near the angle of specular reflection off the sidewalls ($m=3,~4,~5$ in this cartoon). }
\label{fig:catg_path}
\end{figure}

Technically speaking, this translates to an instrument with high grating dispersion and efficiency. The so-called Critical Angle Transmission Grating (CAT grating, \citep{Heilmann2009}) is a promising design to achieve this goal. Fig.~\ref{fig:catg_path} illustrates the dispersion geometry of CAT grating. Although the dispersed X-rays are reflected onto the detector\footnote{Most of the harder X-rays photons are transmitted through the grating.}, interference between the waves coming off the different bars produces a diffraction pattern and the dispersion relation behaves as a transmission grating \citep{Heilmann2022} (cf. Eq.~\ref{eq:disp_tran})
\begin{equation}
\label{eq:disp_cat}
    m \lambda = d(\sin \theta - \sin \beta_m),~
\end{equation}
where $m$ is the spectral order, $d$ the grating period, $\theta$ the incidence angle with respect to the grating bar sidewalls, and $\beta_m$ the $m$th-order diffraction angle. If the incident angle $\theta$ is smaller than a critical angle ($\theta_{\rm crit}$) and the grating sidewalls are adequately smooth, they can act as nanometer-size ``mirrors" to enhance (blaze) the diffraction orders near the direction of specular reflection off the sidewalls. The critical angle $\theta_{\rm crit}$ depends on the wavelength/energy of the incident photon, as well as the reflection index of the grating bar materials \citep{Heilmann2022}. Furthermore, the free-standing tall grating bars should be as thin as possible so that incident photons are not reflected onto the neighboring grating bar \citep{Heilmann2022}.

\begin{figure}
\centering
\includegraphics[width=\hsize, trim={0.cm 0.0cm 0.cm 0.0cm}, clip]{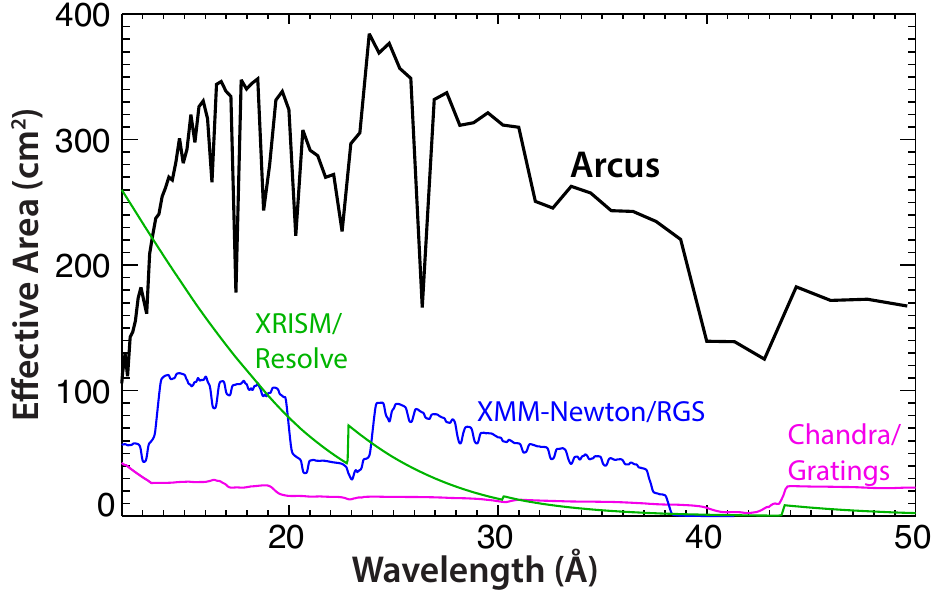}
\caption{Effective area of Arcus (black) in comparison with XRISM/Resolve (green), XMM-Newton/RGS (blue), and Chandra/LEG (purple). Figure reproduced with permission from \cite{Smith2019}, SPIE. }
\label{fig:aeff_arcus}
\end{figure}

\begin{figure}
\centering
\includegraphics[width=\hsize, trim={0.cm 0.0cm 0.cm 0.0cm}, clip]{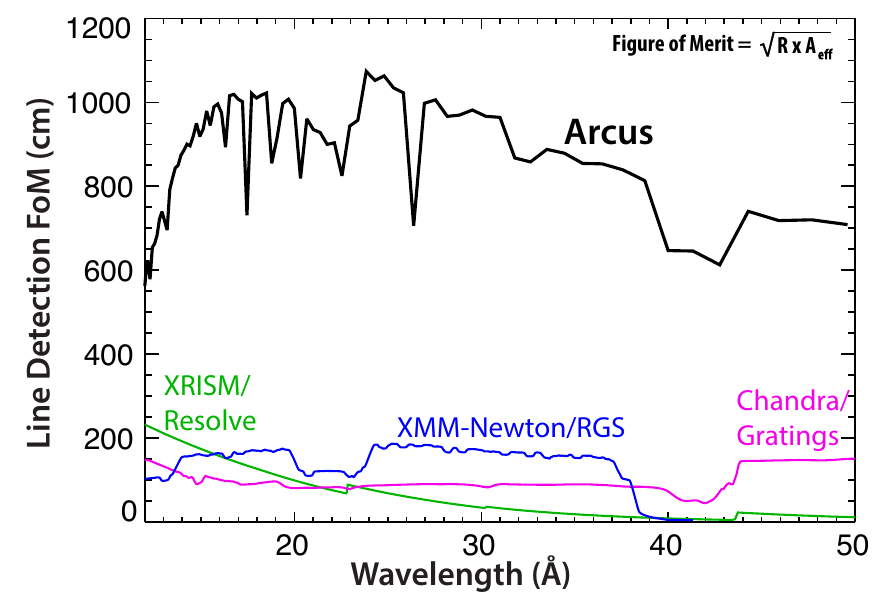}
\caption{The figure of merit (FoM) for detecting weak absorption features. The black, green, blue, and purple curves are Arcus, XRISM/Resolve, XMM-Newton/RGS, and Chandra/LEG, respectively. Figure reproduced with permission from \cite{Smith2019}, SPIE.  }
\label{fig:fom_abs_arcus}
\end{figure}

Arcus will be equipped with CAT gratings to yield a resolving power of $R\sim3800$ over the $10-50$~\AA\ wavelength range \citep{Smith2019}. Thanks to the high grating efficiency, the effective area of Arcus is more than a factor of three larger than RGS aboard XMM-Newton (Fig.~\ref{fig:aeff_arcus}). Combined with its high spectral resolution, Arcus is suitable for detecting weak absorption lines by design (Fig.~\ref{fig:fom_abs_arcus}). Similarly, HiReX will also adopt CAT gratings \citep{Nicastro2021}. It aims to achieve a resolving power $R\gtrsim10^4$ over the wavelength range of $\sim8-124$~\AA\ while having a rather large effective area ($1500~{\rm cm^2}$ at 0.5 keV). 

Another approach to achieving high grating dispersion and efficiency is the so-called off-plan reflection grating \citep{McEntaffer2009}. Unlike conventional (``in-plane") reflection gratings, the off-plan design does not suffer from a strong anti-correlation between dispersion and reflectivity. While CAT gratings require the nanometer-size scaled grating bar to be rather smooth, the off-plan reflection grating is relatively easier to manufacture. However, it requires precise optical alignment (cf. free-standing CAT grating bars). We refer readers to \citep{Paerels2010} and \citep{McEntaffer2009} for technical details.   

\subsection{X-ray spectrometers: diffractive or not}
\label{sct:diff_or_not}
As mentioned earlier, diffractive grating spectrometers can be viewed as constant $\Delta \lambda$ devices so that their resolving power $R=\lambda/\Delta \lambda$ increases with the wavelength of the photon. Micro-calorimeters can be viewed as constant $\Delta E$ devices so that their resolving power ($R=E/\Delta E$) decreases with the wavelength of the photon. At 0.5 keV, a future micro-calorimeter with $\Delta E=0.5$~eV (slightly better than HUBS aims to achieve for its central array) will have a resolving power of $10^3$. At $\lambda=24.797$~\AA\ (i.e., $E=0.5$~keV), Chandra/MEG with $\Delta \lambda=0.023$~\AA\ has already achieved the resolving power of $\sim10^3$ (Table~\ref{tbl:cf_inst}). Hence, diffractive grating spectrometers will still be the leading designs in terms of resolving power in the soft X-ray band.

The large resolving power of diffractive grating spectrometers comes with a price though. These devices are optimized for point sources. The energy resolution of the device is degraded for extended sources because it is hard to disentangle the dispersion effect from the different optical paths along which the photons of an extended source are dispersed. The effective area of diffractive grating spectrometers is much smaller than that of non-diffractive ones. Furthermore, micro-calorimeters with thousands of pixels (integral field units) are more efficient in obtaining a large number of high-resolution X-ray spectra in one single observation. 

\section{XMM-Newton Reflection Grating Spectrometer}
\label{sct:xmm_rgs}
In the following, we focus on the XMM-Newton Reflection Grating Spectrometer. XMM-Newton \citep{Jansen2001} is the second cornerstone project of the ESA's Horizon 2000 Science Programme. It has a set of three X-ray CCD cameras, comprising the European Photon Imaging Camera (EPIC). Two of them are MOS (Metal Oxide Semiconductor) \citep{Turner2001}. The third one is a pn device \citep{{Struder2001}}. All three CCD imaging spectrometers ($\sim30$~arcmin field-of-view and $\sim6$~arcsec angular resolution) are non-diffractive spectrometers. The resolving power of EPIC is $R=E/\Delta E\sim20-50$. 

\begin{figure}
\centering
\includegraphics[width=.7\hsize, trim={0.5cm 0.0cm 1.0cm 0.5cm}, clip]{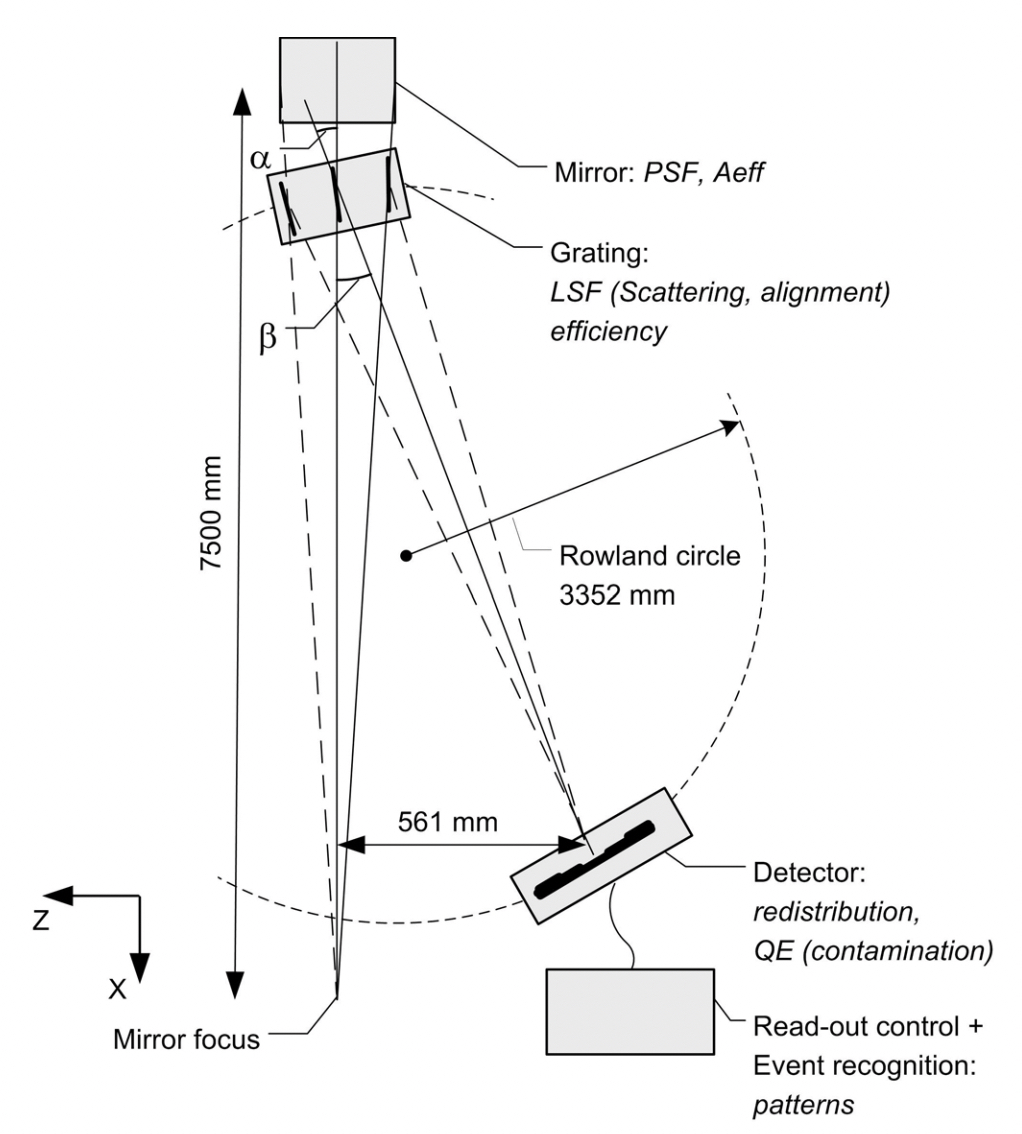}
\caption{Optical design of XMM-Newton Reflection Grating Spectrometer (not to scale). X-rays enter from the top. About half of the soft X-ray photons arrive at the mirror focus. The other half was reflected onto the RGS Focal Camera (RFC). Figure reproduced with permission from \cite{deVries2015}, A\&A. }
\label{fig:rgs_path}
\end{figure}

Behind each MOS camera is a Reflection Grating Array (RGA, Fig.~\ref{fig:rgs_path}). About half of the incident soft X-ray photons are reflected onto the RGS Focal Camera (RFC). The latter consists of 9 back-illuminated MOS CCDs (similar to those of the EPIC/MOS) in a row along the dispersion direction. 

\begin{figure}
\centering
\includegraphics[width=\hsize, trim={0.5cm 0.0cm 0.5cm 0.5cm}, clip]{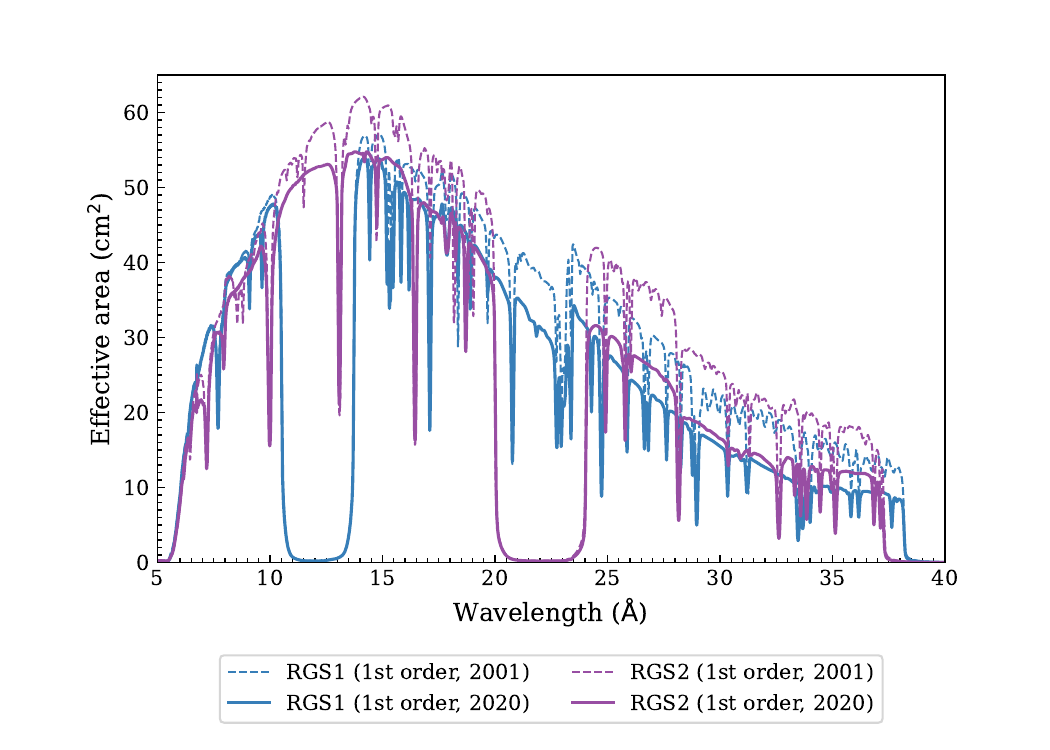}
\caption{Effective area of RGS1 (blue) and RGS2 (purple). The solid and dashed curves are the effective area in 2001 and 2020, respectively. Broad and narrow dips are instrument effects (see text for details). }
\label{fig:plot_aeff_rgs}
\end{figure}

Fig.~\ref{fig:plot_aeff_rgs} shows the effective area of both RGS instruments for the first order. Dozens of narrow dips are caused by CCD gaps and hot pixels \citep{deVries2015}. The two broad troughs $\sim11-14$~\AA\ and $\sim20-24$~\AA\ are due to failures of CCD \#7 of RGS1 and CCD \#4 of RGS2 at the beginning of the mission (and are therefore not present for observations earlier than 2000-09-02) \citep{deVries2015}. Furthermore, over the years, a gradual decline in the effective area for the entire wavelength range can be noticed.

\begin{figure}
\centering
\includegraphics[width=\hsize, trim={0.5cm 0.0cm 0.5cm 0.5cm}, clip]{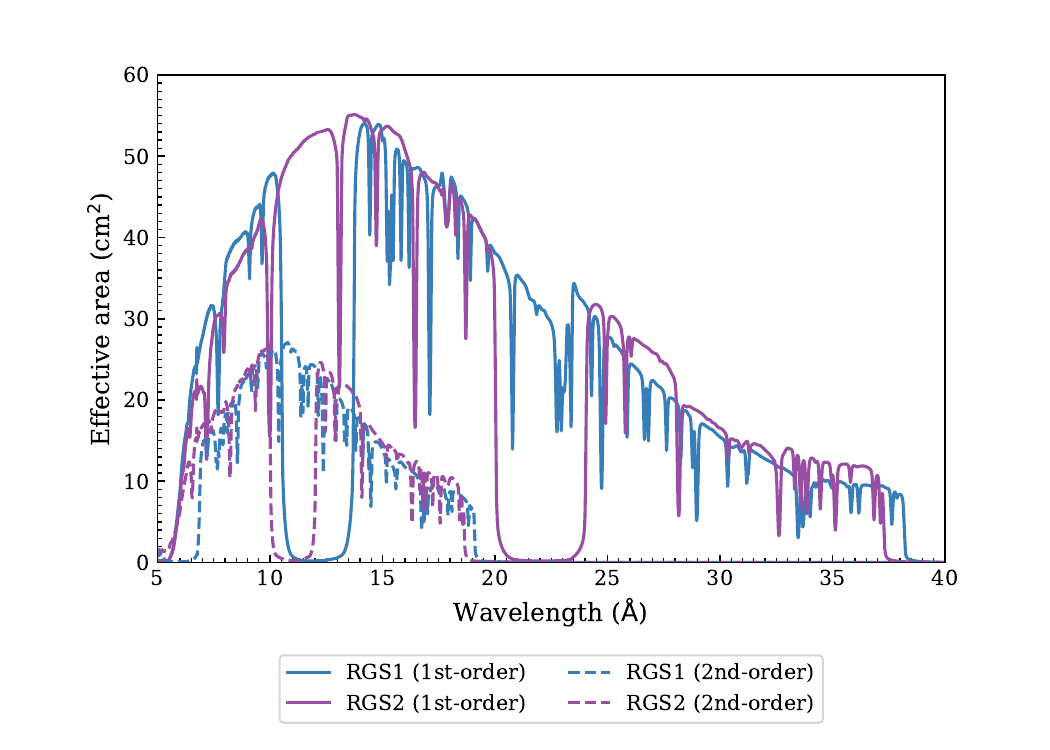}
\caption{Effective area of the RGS spectra are shown in solid and dashed lines, respectively. The 1st- and 2nd-order spectra are shown in blue and purple, respectively. These curves are based on Obs.ID$=$0791980501 targeting HR\,1099 (observed on 2020-02-25). }
\label{fig:plot_aeff_rgs_cford}
\end{figure}

The second-order RGS spectra ($5-19$~\AA\ or $0.65-2.5$~keV) overlap with the first-order spectra ($5-38$~\AA\ or $0.33-2.5$~keV). The two spectra can be distinguished with the intrinsic CCD energy resolution of RFC. Although the second-order RGS spectra have a higher spectral resolution, they are less frequently used (e.g., Pinto et al. 2016 \citep{Pinto2016b}). This is mainly due to the relatively small effective area of the second-order (Fig.~\ref{fig:plot_aeff_rgs_cford}).

\section{RGS data reduction}
\label{sct:rgs_data}
Here, we illustrate how to access RGS data (Section~\ref{sct:rgs_data_access}) and how to reduce RGS (imaging, timing, and spectral) data for the benefit of new learners. For the latter, we include both a general recipe in Sections~\ref{sct:rgs_dr_general} and guidance to handle special cases (Section~\ref{sct:rgs_dr_special}) 

\subsection{RGS data access}
\label{sct:rgs_data_access}
\begin{figure}
\centering
\includegraphics[width=\hsize, trim={0.5cm 0.0cm 0.cm 0.5cm}, clip]{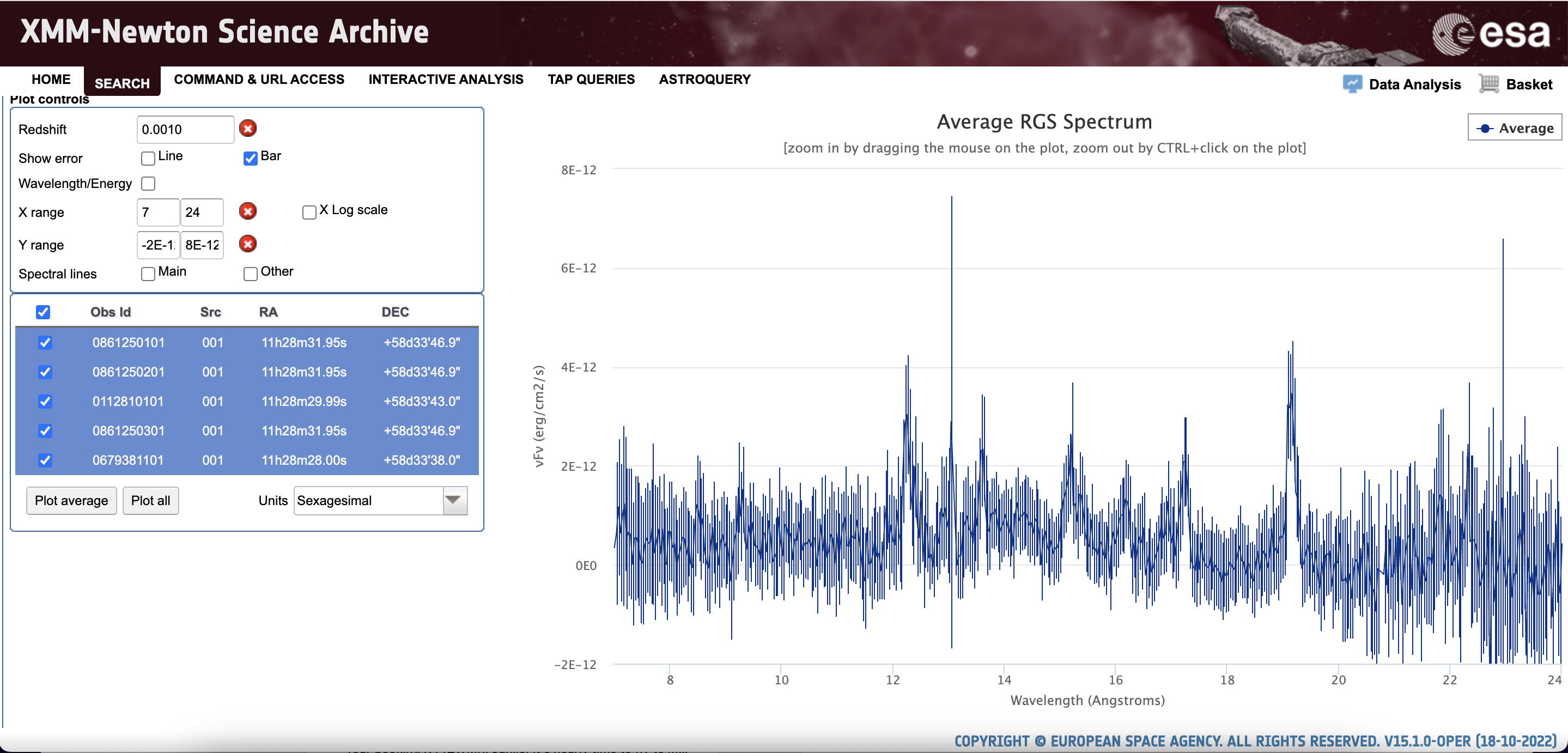}
\caption{XMM-Newton Science Archive (XSA) RGS spectrum plotting function. The averaged RGS spectrum is shown and the rest-frame wavelength is shown because the redshift of the observing target (NGC\,3690/Arp\,299) is provided. }
\label{fig:xsa_rgs}
\end{figure}

\begin{figure}
\centering
\includegraphics[width=\hsize, trim={0.5cm 0.0cm 0.5cm 0.cm}, clip]{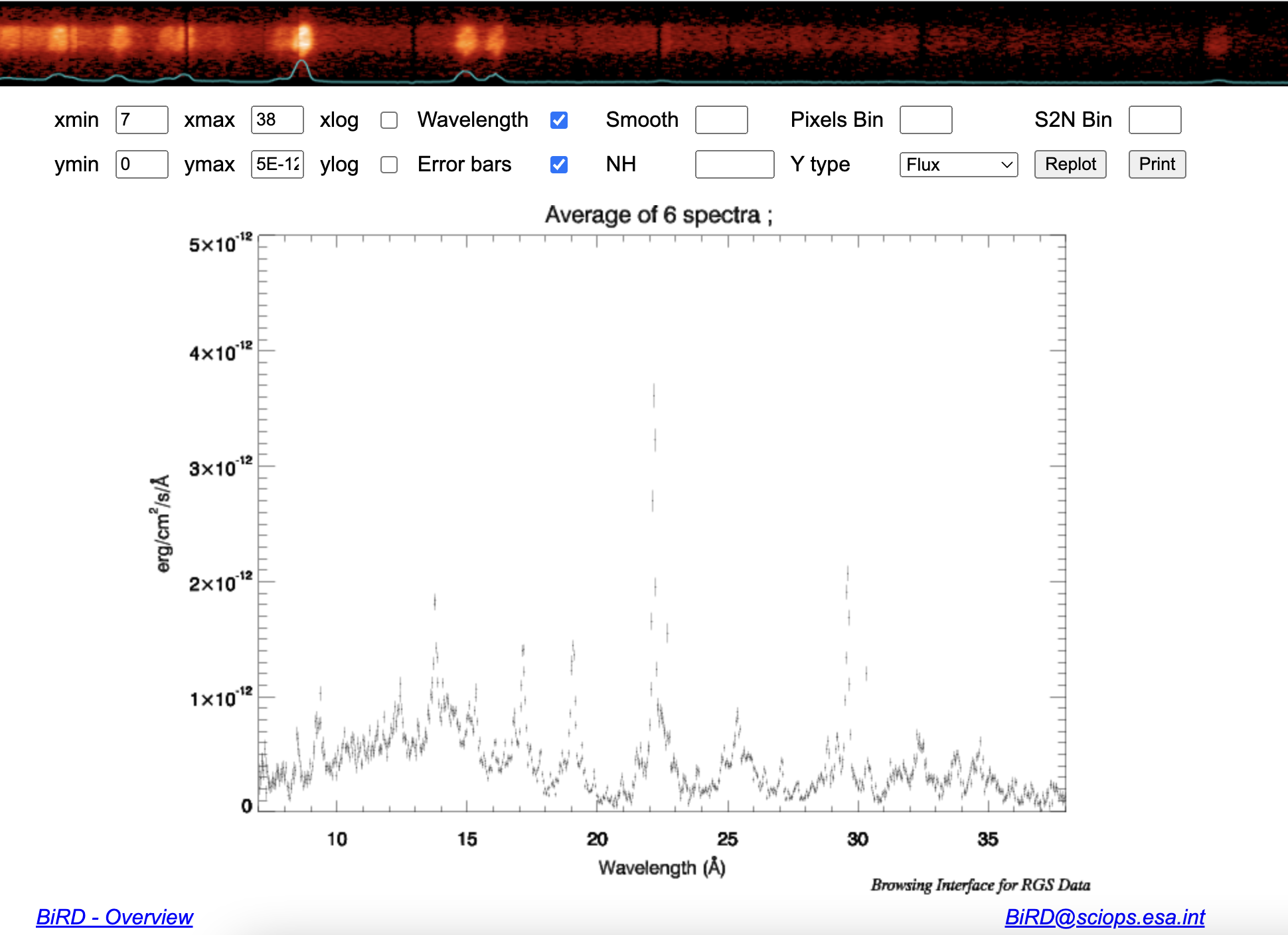}
\caption{XMM-Newton Browsing Interface for RGS Data. The averaged RGS spectrum is shown for NGC\,1068 in the observed frame. The list of observation IDs is shown as part of the query results, along with RGS and EPIC images. }
\label{fig:bird_rgs}
\end{figure}

To simply view RGS data for a certain object, one can use  either the XMM-Newton Science Archive (XSA) \footnote{http://nxsa.esac.esa.int/nxsa-web/\#search} or Browsing Interface for RGS Data (BiRD\footnote{https://xmmweb.esac.esa.int/BiRD/}). The former visualize interactive RGS flux spectra (Fig.~\ref{fig:xsa_rgs}). The latter provides RGS flux spectra as well as RGS and EPIC images (Fig.~\ref{fig:bird_rgs}). 

XSA is also the main channel to access the full XMM-Newton data products. Users can download Observation Data Files (ODF) via a web browser\footnote{http://nxsa.esac.esa.int/nxsa-web/\#search} and the Archive Inter Operability (AIO) command line client\footnote{http://nxsa.esac.esa.int/nxsa-web/\#aio}. After downloading the ODF data, Science Analysis System (SAS) will be used to reduce the data. Note that, for archival observations, it is possible to reduce the data via the Remote Interface for Science Analysis (RISA)\footnote{http://nxsa.esac.esa.int/nxsa-web/\#risa\_introduction} server without downloading data and software. 

In the following, we explain how to reduce data with SAS. This assumes that SAS and its dependency packages have been successfully installed and initialized\footnote{https://www.cosmos.esa.int/web/xmm-newton/sas-installation}. Some useful tips can be found in the XMM-Newton ABC Guide\footnote{https://heasarc.gsfc.nasa.gov/docs/xmm/abc/abc.html}.

\subsection{General guide for RGS data reduction using SAS}
\label{sct:rgs_dr_general}
We provide a typical recipe to generate RGS imaging, timing, and spectral data products. We mainly use Obs.ID$=$0791980501 observed on 2020-02-25 with a duration of $\sim$50 ks. The target HR\,1099 (a star) is a point-like source. 


\subsubsection{Getting started}
\label{sct:rgs_dr_prep}
Download the ODF data of Obs.ID$=$0791980501 (for HR\,1099) to the main working directory. Users have the freedom to structure the main working directory according to their preferences. Here, we structure the main working directory as follows\footnote{Bash shell commands are provided throughout the document. C-shell commands can certainly be used though.}:
\begin{svgraybox}
\noindent user\$ \# Lines starting with a hash-tag is for comments 

\noindent user\$ \# Define the dir\_main variable

\noindent user\$ dir\_main=/path/to/the/main/working/directory 

\noindent user\$ cd \$\{dir\_main\}

\noindent user\$ \# Create the following sub-directories

\noindent user\$ mkdir odf

\noindent user\$ mkdir epic

\noindent user\$ mkdir rgs

\noindent user\$ ls 

\noindent \textcolor{blue}{odf epic rgs}

\noindent user\$ cd \$\{dir\_main\}/odf

\noindent user\$ \# Define the obsid variable

\noindent user\$ obsid=0791980501 

\noindent user\$ curl -o \$\{obsid\}\_odf.tar "http://nxsa.esac.esa.int/nxsa-sl/servlet/data-action-aio?obsno=\${obsid}\&level=ODF"

\end{svgraybox}

It might take a few minutes to complete the downloading process, depending on the data file size, the network speed, and so forth. As mentioned earlier, one can also download the ODF data via the XSA website. Subsequently, we unpack the ODF data and set the SAS\_ODF environment variable. 

\begin{svgraybox}
\noindent user\$ tar -xvf \${obsid}\_odf.tar

\noindent user\$ tar -xf *.TAR


\noindent user\$ SAS\_ODF=\$\{dir\_main\}/\$\{obsid\}/odf
 
\noindent user\$ export SAS\_ODF



\end{svgraybox}

Before leaving the \${dir\_main}/odf directory, we have to create the ccf.cif file, which is an index file of Current Calibration Files (CCF). The SAS\_CCF environment variable should refer to the ccf.cif file. 

\begin{svgraybox}

\noindent user\$ cifbuild


\noindent user\$ SAS\_CCF=\$\{dir\_main\}/\$\{obsid\}/odf/ccf.cif
 
\noindent user\$ export SAS\_CCF



\end{svgraybox}

Run the \textit{odfingest} task to extend the ODF summary file, whose file name matches the pattern of *SUM.SAS, with data extracted from the instrument housekeeping data files and the calibration database. Once this task is finished, users have to update the SAS\_ODF environment to the extended ODF summary file. 

\begin{svgraybox}
\noindent user\$ odfingest 


\noindent user\$ SAS\_ODF=\$(ls \$\{dir\_main\}/\$\{obsid\}/odf/*SUM.SAS)

\noindent user\$ export SAS\_ODF



\end{svgraybox}

\subsubsection{Running the RGS data reduction pipeline \textit{rgsproc}}
\label{sct:rgs_proc1}
Although the pipeline can be run directly using all default parameters, users are strongly encouraged:
\begin{itemize}
    \item To set the source coordinates in degrees by setting the parameter withsrc to yes, as well as specifying srclabel, srcra, and srcdec parameters. This is crucial because the source coordinates have a profound influence on the accuracy of the wavelength scale as recorded in the response file generated by \textit{rgsproc}.
    \item To activate the RGS effective area correction by setting the parameter witheffectiveareacorrection to yes. This correction is based on the careful analysis after \citet{Kaastra2018}. For SAS v19.0 (released at the end of 2020) and later versions, witheffectiveareacorrection is set to yes by default. 
    \item To activate the RGS background model spectrum by setting the parameter withbackgroundmodel to yes. This is particularly useful if the source extends a large fraction of the RGS field or view or if the local background has poor statistics. For bright point-like sources, the local background is sufficient. 
\end{itemize}

\begin{svgraybox}
\noindent user\$ cd \$\{dir\_main\}/rgs

\noindent user\$ \# HR\,1099 sky coordinates from Simbad/NED

\noindent user\$ ra\_deg=54.1970

\noindent user\$ dec\_deg=0.5878

\noindent user\$ \# The source label (srclabel) is defined by the user but should avoid PROPOSAL and ONAXIS. 

\noindent user\$ rgsproc withsrc=yes srclabel=USER srcra=\$\{ra\_deg\} 

srcdec=\$\{dec\_deg\} witheffectiveareacorrection=yes  

withbackgroundmodel=yes 

\end{svgraybox}

It might take a few minutes to generate all the products. These output files are named following the Pipeline Processing Subsystem (PPS) file name convention in the current working directory (\${dir\_main}/rgs in this example). 

\subsubsection{Extracting RGS images}
\label{sct:rgs_img}
\begin{figure}
\centering
\includegraphics[width=\hsize, trim={0.5cm 0.0cm 0.5cm 0.5cm}, clip]{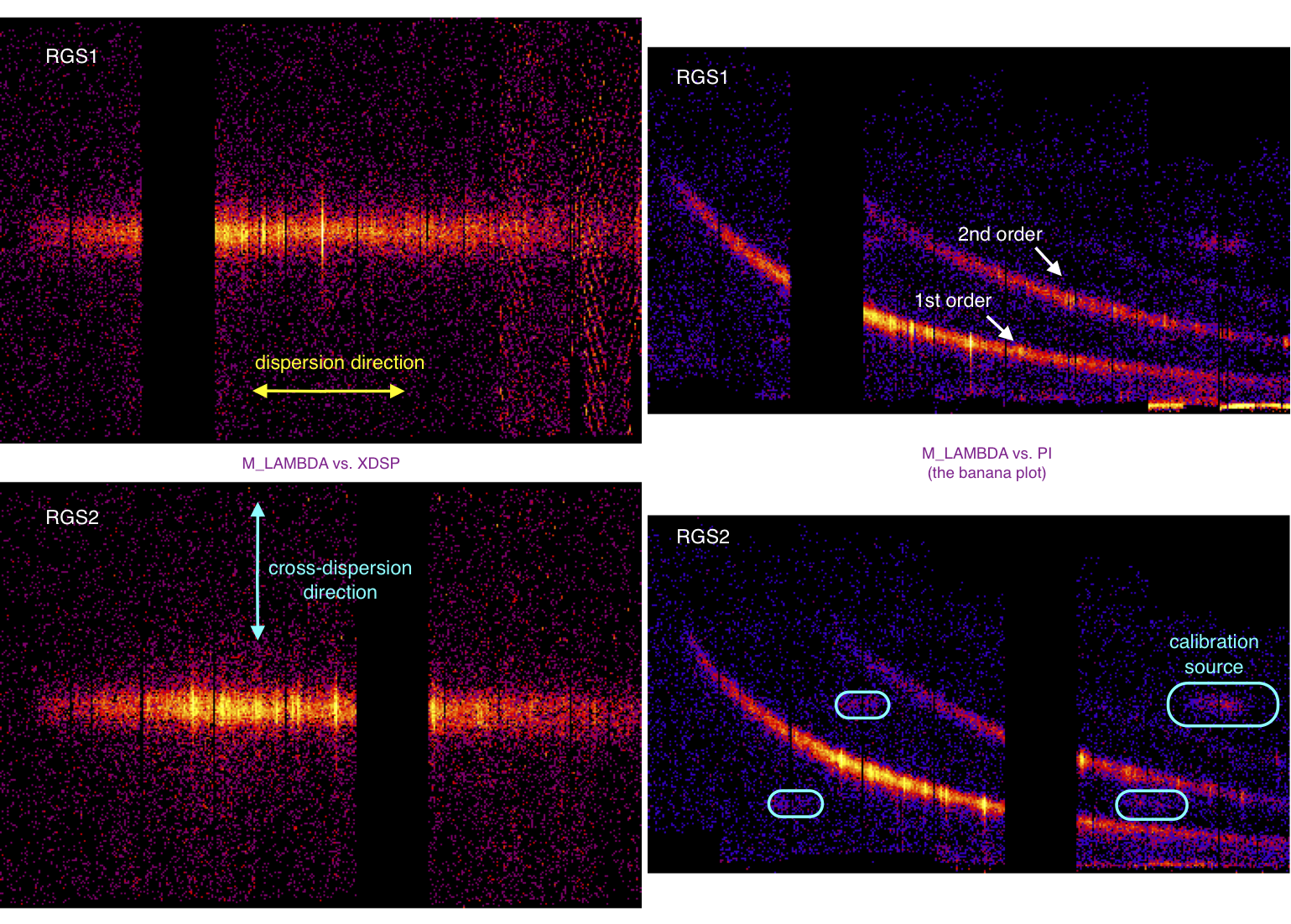}
\caption{RGS images for Obs.ID$=$0791980501 (HR\,1099). The left two panels are displayed in the M\_LAMBDA vs. XDSP\_CORR parameter space while the right two panels are displayed in the M\_LAMBDA vs. PI parameter space. The upper and lower two panels are RGS1 and RGS2, respectively. Failure of CCD \#7 of RGS1 and CCD \#4 of RGS2 are clearly visible. }
\label{fig:0791980501_rgs_img}
\end{figure}

There are two types of RGS images can be extracted from the event file produced by \textit{rgsproc}. The first type is in the M\_LAMBDA vs. XDSP\_CORR  parameter space while the second type is in the M\_LAMBDA vs. PI parameter space. M\_LAMBDA is $m\lambda$, where $m=1,~2$ is the spectral order and $\lambda$ is the wavelength of the dispersed photon. XDSP\_CORR reflects the extension of the source in the cross dispersion direction. PI is related to the energy of the dispersed photon. Both types of images are extracted in a similar way. The event file has the following name convention: P\$\{obsid\}R\$\{rgsid\}\$\{expid\}EVENLI0000.FIT, where \$\{rgsid\} is either 1 (for RGS1) or 2 (for RGS2), \$\{expid\} is the exposure ID of the instrument. The exposure ID starts with either S (for scheduled observations) or U (for unscheduled observations), followed by a three-digit number. For obsid$=$0791980501, we have expid$=$S004 for RGS1 and expid$=$S005 for RGS2. Fig.~\ref{fig:0791980501_rgs_img} shows the RGS images for obsid$=$0791980501. 

\begin{svgraybox} 
\noindent user\$ rgsid=1

\noindent user\$ expid=S004

\noindent user\$ lis\_evt=P\$\{obsid\}R\$\{rgsid\}\$\{expid\}EVENLI0000.FIT

\noindent user\$ \# The output image file name (imageset) is defined by the user. 

\noindent user\$ evselect table=\$\{lis\_evt\}:EVENTS 

imageset=img\_rgs1\_xdsp\_dsp.fits xcolumn='M\_LAMBDA' 

ycolumn='XDSP\_CORR'

\noindent user\$ \# Next, we extract the M\_LAMBDA vs. PI plot 

\noindent user\$ \# and specify the size of the images (600 pixels $\times$ 600 pixels), 

\noindent user\$ \# which can be defined by the user. 

\noindent user\$ evselect table=\$\{lis\_evt\}:EVENTS 

imageset=img\_rgs1\_pi\_dsp.fits xcolumn='M\_LAMBDA' 

ycolumn='PI' imagebinning=imageSize ximagesize=900 

yimagesize=600 
\end{svgraybox}

\begin{figure}
\centering
\includegraphics[width=\hsize, trim={0.5cm 0.0cm 0.5cm 0.5cm}, clip]{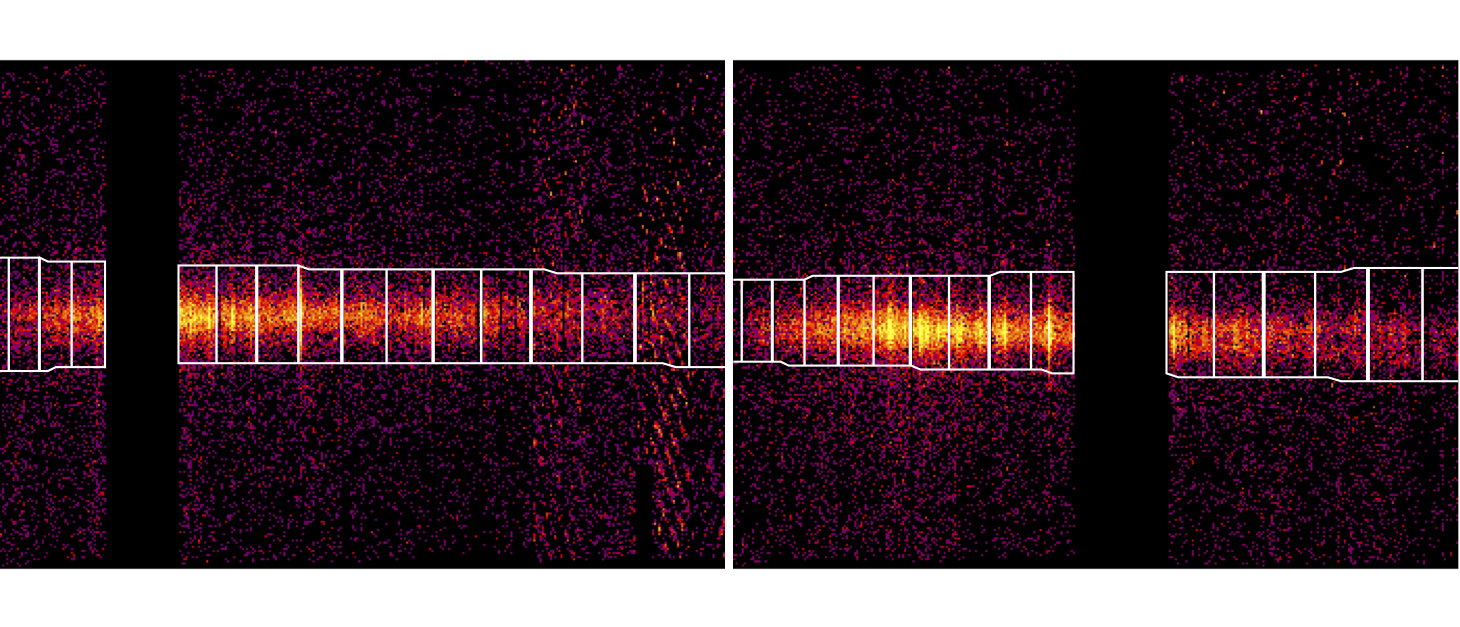}
\caption{RGS source extraction region (90\% PSF along the cross-dispersion direction) for Obs.ID$=$0791980501 (HR\,1099). The left and right panels are RGS1 and RGS2, respectively. }
\label{fig:0791980501_rgs_src_reg}
\end{figure}

In Section~\ref{sct:rgs_proc1}, we used the default extraction region for the source along the cross-dispersion direction (i.e., xpsfincl$=90$ for \textit{rgsproc} by default), which is 90\% of the telescope point spread function (PSF). This can be visualized by first generating the extract region from the source list file via the \textit{cxctods9} task and then loading the region file along with the M\_LAMBDA vs. XDSP\_CORR image on ds9 (Fig.~\ref{fig:0791980501_rgs_src_reg}). 90\%, 95\%, 98\%, and 99\% of the telescope PSF corresponds to  $\sim0.8$~arcmin, $\sim1.4$~arcmin, $\sim2.8$~arcmin, and $\sim3.4$~arcmin, respectively, along the cross dispersion direction. 

\begin{svgraybox} 
\noindent user\$ rgsid=1

\noindent user\$ expid=S004

\noindent user\$ lis\_src=P\$\{obsid\}R\$\{rgsid\}\$\{expid\}SRCLI\_0000.FIT

\noindent user\$ \# The source ID (srcid) can be found in the source list file \$\{lis\_src\}

\noindent user\$ srcid=3

\noindent user\$ \# The output region file name is defined by the user. 

\noindent user\$ cxctods9 table=\$\{lis\_src\}:RGS\$\{rgsid\}\_SRC\$\{srcid\}\_SPATIAL 

regtype=linear -V 0 $>$ rgs\$\{rgsid\}\_src.reg

\end{svgraybox}

\subsubsection{Extracting RGS lightcurves}
\begin{figure}
\centering
\includegraphics[width=\hsize, trim={0.5cm 0.0cm 0.5cm 0.5cm}, clip]{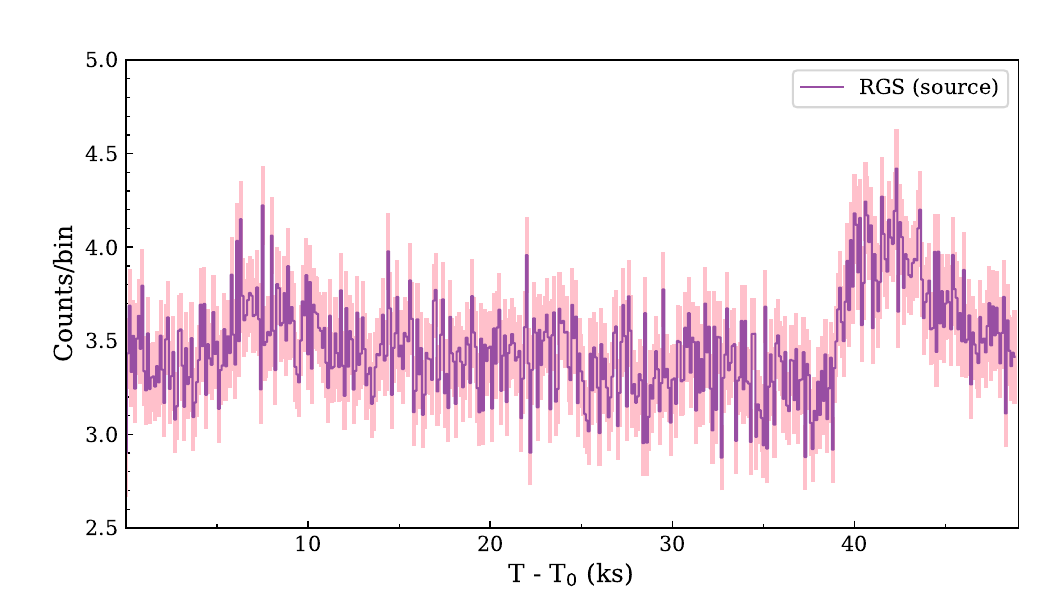}
\caption{RGS source lightcurve for Obs.ID$=$0791980501 (HR\,1099).}
\label{fig:0791980501_rgs_src_ltc}
\end{figure}

To get a barycentric corrected and background subtracted RGS lightcurve, one should use the \textit{rgslccorr} task. In the following example, we create an RGS1 and RGS2 combined, 1st order barycentric-corrected and background-subtracted light curve with a time bin size of 100 s, for the 3rd source in the source list (Fig.~\ref{fig:0791980501_rgs_src_ltc}):

\begin{svgraybox} 
\noindent user\$ \# Time bin size (100 s) is defined by the user

\noindent user\$ \# The output file name (ltc\_rgs\_src.fits) is also defined by the user

\noindent user\$ rgslccorr evlist="P\$\{obsid\}R1S004EVENLI0000.FIT 

P\$\{obsid\}R2S005EVENLI0000.FIT"

srclist="P\$\{obsid\}R1S004SRCLI\_0000.FIT 

P\$\{obsid\}R2S005SRCLI\_0000.FIT"

timebinsize=100 orders='1' sourceid=3 

outputsrcfilename=ltc\_rgs\_src.fits
\end{svgraybox}

If users would like to study the RGS imaging and spectral data in the last $\sim10$~ks of this observation, they can generate a Good Time Interval (GTI) file using either \textit{gtibuild} or \textit{tabgtigen}. The former requires a text file in the ASCII format with three columns and one to many rows. The first two columns are the start and end time (in seconds) for an interval, while the last column is either ``+" (to keep in the analysis) or ``-" (to discard).  

\begin{svgraybox} 
\noindent user\$ \# The input file name (gti.txt) can be defined by the user

\noindent user\$ cat gti.txt 

\noindent 699021517.798482 699030397.999542 +

\noindent user\$ gtibuild file=gti.txt table=gti.fits
\end{svgraybox}

Alternatively, the \textit{tabgtigen} task can be used as follows: 

\begin{svgraybox} 
\noindent user\$ \# The input file name (gti.fits) can be defined by the user

\noindent user\$ tabgtigen table=ltc\_rgs\_src.fits  gtiset=gti.fits timecolumn=TIME 

expression=`(TIME in [6.990215E8:6.990304E8])'
\end{svgraybox}      

We caution that the XMM-Newton is susceptible to background flares. In Section~\ref{sct:rgs_proc1}, we generate imaging, timing, and spectra products using all the exposure of the observation. While RGS suffers less from the background flares than EPIC, it is still highly recommended to examine the background lightcurve. The RGS background lightcurve is extracted from CCD \#9 for two reasons: (1) photons arriving at CCD \#9 (with $\lambda\sim5-7$~\AA) are more sensitive to background flares; (2) CCD \#9 records the least source photons due to its location close to the optical axis. Furthermore, we extract the background lightcurve in a region away from the source (along the cross-dispersion direction) as follows:

\begin{svgraybox} 
\noindent user\$ \# Time bin size (100 s) is defined by the user

\noindent user\$ \# The output file name (ltc\_rgs1\_bkg.fits) is also defined by the user

\noindent user\$ evselect table=\$\{lis\_evt\} timebinsize=100 rateset=ltc\_rgs1\_bkg.fits 

makeratecolumn=yes maketimecolumn=yes 

expression="(CCDNR==9)\&\&(REGION(\$\{lis\_src\}:

RGS\$\{rgsid\}\_BACKGROUND,M\_LAMBDA,XDSP\_CORR))"

\end{svgraybox}

\begin{figure}
\centering
\includegraphics[width=\hsize, trim={0.5cm 0.0cm 0.5cm 0.5cm}, clip]{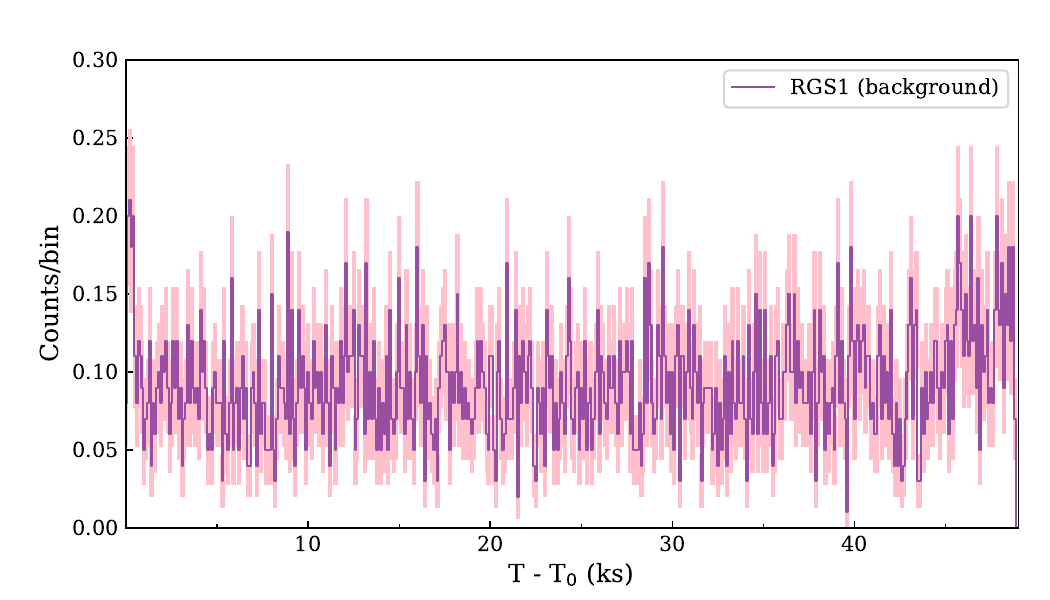}
\caption{RGS1 background lightcurve for Obs.ID$=$0791980501 (HR\,1099).}
\label{fig:0791980501_rgs_bkg_ltc}
\end{figure}

As shown in Fig.~\ref{fig:0791980501_rgs_bkg_ltc}, this exemplary observation does not suffer from background flares. Nonetheless, the total exposure can be significantly reduced for e.g., Obs.ID$=0116710901$ (HR\,1099). In that case, users can create a GTI with a threshold count rate (all those below this threshold are kept as GTI) using the \textit{tabgtigen} task. Caution that the threshold can vary significantly for different observations of the same target, let alone different targets. 

\begin{svgraybox} 
\noindent user\$ \# The threshold count rate can be defined by the user. 

\noindent user\$ tabgtigen table=ltc\_rgs\_src.fits  gtiset=gti.fits timecolumn=TIME 

expression="(RATE $<=$ 4.049)"
\end{svgraybox} 

\begin{figure}
\centering
\includegraphics[width=\hsize, trim={0.1cm 0.0cm 0.1cm 0.5cm}, clip]{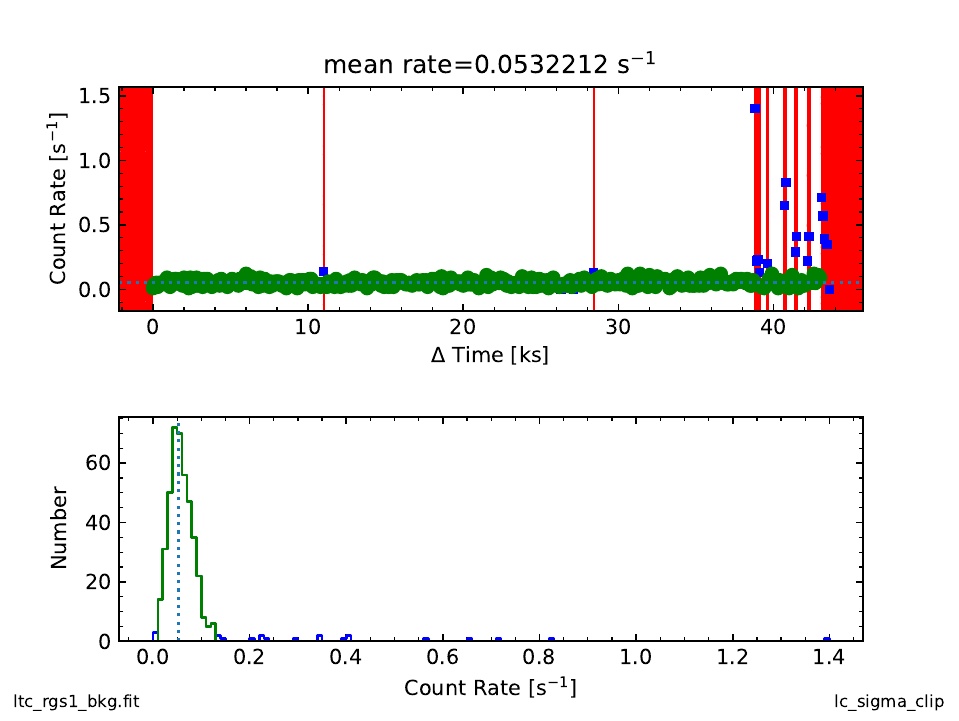}
\caption{RGS1 background lightcurve for Obs.ID$=$0134540101 (HR\,1099). Time intervals impacted by flares are identified using the sigma clipping method via the \textit{deflare} task provided by CIAO. The threshold used in this example is $3\sigma$. The shaded area in red also includes time intervals with zero count rates. Good time intervals are those highlighted in green.}
\label{fig:0134540101_rgs_bkg_ltc}
\end{figure}

Alternatively, users can determine the GTI using the so-called sigma clipping method via the \textit{deflare} task provided by CIAO (Chandra Interactive Analysis of Observations)\footnote{https://cxc.cfa.harvard.edu/ciao/ahelp/deflare.html}. Figure~\ref{fig:0134540101_rgs_bkg_ltc} shows an example of such an application (see also \citep{Mao2021}). In order to apply the outcome of \textit{deflare} to the RGS data reduction with SAS, one needs to delete the first extension of the \textit{deflare} output file. This can be realized using the \textit{fdelhdu}\footnote{https://heasarc.gsfc.nasa.gov/lheasoft/ftools/fhelp/fdelhdu.html} or \textit{ftdelhdu}\footnote{https://heasarc.gsfc.nasa.gov/lheasoft/ftools/headas/ftdelhdu.html} task available from the Heasoft package.

\begin{svgraybox} 
\noindent user\$ \# Assuming gti.fits is the output file of deflare 

\noindent user\$ fdelhdu gti.fits+1 N Y

\noindent user\$ \# Use ftdelhdu as an alternative

\noindent user\$ ftdelhdu infile=''gti.fits[FILTER]" outfile=gti.fits clobber=1
\end{svgraybox}

To apply the GTI filtering, we need to run \textit{rgsproc} again. The first two steps (events and angles) of \textit{rgsproc} might be skipped by specifying the entry and final stages.

\begin{svgraybox}
\noindent user\$ rgsproc auxgtitables=gti.fits entrystage=3:filter finalstage=5:fluxing

witheffectiveareacorrection=yes  
 withbackgroundmodel=yes 
\end{svgraybox}

\subsubsection{Extracting RGS spectra}
The pipeline products of \textit{rgsproc} contain RGS spectra files for both instruments (rgsid$=1$ or 2) and both spectral orders. Note that for RGS, the response matrix file (rmf) and ancillary response file (arf) are combined into one response file. 
\begin{itemize}
    \item P\$\{obsid\}R\$\{rgsid\}\$\{expid\}BGSPEC100\$\{srcid\}.FIT: 1st-order local background spectrum file
    \item P\$\{obsid\}R\$\{rgsid\}\$\{expid\}BGSPEC200\$\{srcid\}.FIT: 2nd-order local background spectrum file
    \item P\$\{obsid\}R\$\{rgsid\}\$\{expid\}MBSPEC1000.FIT: 1st-order model background spectrum
    \item P\$\{obsid\}R\$\{rgsid\}\$\{expid\}MBSPEC2000.FIT: 2nd-order model background spectrum
    \item P\$\{obsid\}R\$\{rgsid\}\$\{expid\}RSPMAT100\$\{srcid\}.FIT: 1st-order response file
    \item P\$\{obsid\}R\$\{rgsid\}\$\{expid\}RSPMAT200\$\{srcid\}.FIT: 2nd-order response file
    \item P\$\{obsid\}R\$\{rgsid\}\$\{expid\}SRSPEC100\$\{srcid\}.FIT: 1st-order source spectral file
    \item P\$\{obsid\}R\$\{rgsid\}\$\{expid\}SRSPEC200\$\{srcid\}.FIT: 2nd-order source spectral file
\end{itemize}

To combine RGS1 and RGS2 spectra of the same spectral order, one can take advantage of the \textit{rgscombine} task. This applies to either one observation or multiple observations. The example below combines the first-order RGS1 and RGS2 spectra (with the model background) for Obs.ID$=0791980501$:

\begin{svgraybox}
\noindent user\$ \# First, create some lists of files to be combined (src.lis, rsp.lis, bkg.lis)

\noindent user\$ cat src.lis

\noindent P0791980501R1S004SRSPEC1003.FIT~~ 
P0791980501R2S005SRSPEC1003.FIT

\noindent user\$ cat rsp.lis

\noindent P0791980501R1S004RSPMAT1003.FIT~~ 
P0791980501R2S005RSPMAT1003.FIT

\noindent user\$ cat bkg.lis

\noindent P0791980501R1S004MBSPEC1000.FIT~~ 
P0791980501R2S005MBSPEC1000.FIT

\noindent user\$ \# Output files (filepha, filermf, and filebkg) are defined by the user

\noindent user\$ rgscombine pha="\$(cat src.lis)" rmf="\$(cat rsp.lis)" 

bkg="\$(cat bkg.lis)" filepha=rgs\_o1\_src.fits 

filermf=rgs\_o1.rsp ilebkg=rgs\_o1\_bkg.fits

\end{svgraybox}

\begin{figure}
\centering
\includegraphics[width=\hsize, trim={0.5cm 0.0cm 0.5cm 0.5cm}, clip]{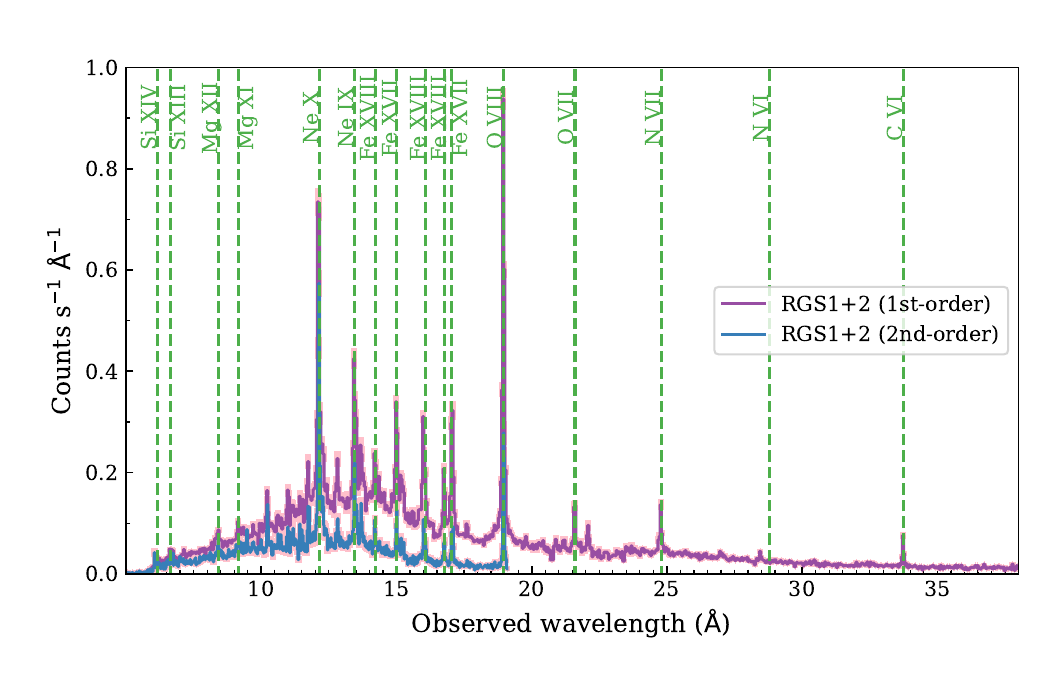}
\caption{RGS combined and folded spectrum for Obs.ID$=$0791980501 (HR\,1099). The 1st- and 2nd-order spectra are shown in purple and blue, respectively. Uncertainties are shown in pink. Some key diagnostic lines are labeled in green.  }
\label{fig:plot_spec_rgs_fold}
\end{figure}

Fig.~\ref{fig:plot_spec_rgs_fold} compares the 1st- and 2nd-order combined and folded spectra of Obs.ID$=$0791980501 (HR\,1099). Due to the smaller effective area (Fig.~\ref{fig:plot_aeff_rgs_cford}), the 2nd-order spectrum is lower than the 1st-order in this plot. 

RGS fluxed spectra should also be available among the \textit{rgsproc} pipeline products. They can also be generated with the \textit{rgsfluxer} task: 

\begin{svgraybox}
\noindent user\$ \# Output file (rgs\_o1\_flux.fits) is defined by the user

\noindent user\$ cat src.lis

\noindent P0791980501R1S004SRSPEC1003.FIT~~ 
P0791980501R2S005SRSPEC1003.FIT

\noindent user\$ cat rsp.lis

\noindent P0791980501R1S004RSPMAT1003.FIT~~ 
P0791980501R2S005RSPMAT1003.FIT

\noindent user\$ \# User can include the background files via the keyword agrument ``bkg"

\noindent user\$ \# We skip the background setting here for simplicity. 

\noindent user\$ rgsfluxer pha="\$(cat src.lis)" rmf="\$(cat rsp.lis)" file=rgs\_o1\_flux.fits
\end{svgraybox}

\begin{figure}
\centering
\includegraphics[width=\hsize, trim={0.2cm 0.0cm 0.5cm 0.5cm}, clip]{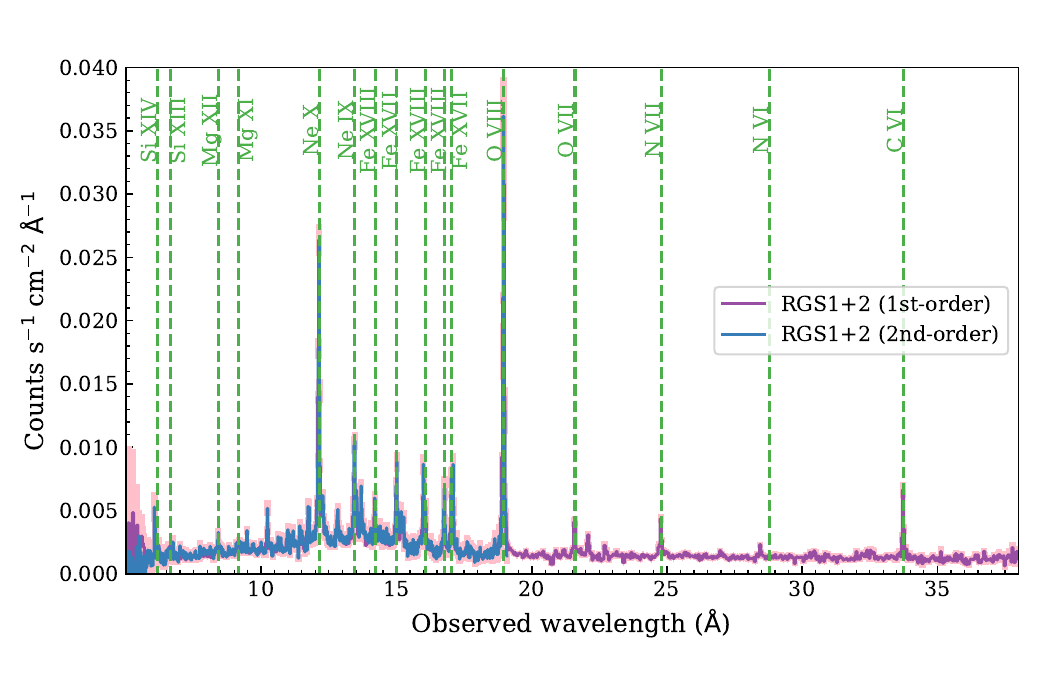}
\caption{Combined and fluxed (i.e., unfolded) RGS spectra for Obs.ID$=$0791980501 (HR\,1099). The 1st- and 2nd-order spectra are shown in purple and blue, respectively. Uncertainties are shown in pink. Some key diagnostic lines are labeled in green. The 1st- and 2nd- order spectra should agree in flux unless the target is piled up. }
\label{fig:plot_spec_rgs_unfold}
\end{figure}

In Fig.~\ref{fig:plot_spec_rgs_unfold}, we compare the combined (RGS1 and RGS2 coadded for the same spectral order) and fluxed RGS spectra for Obs.ID$=$0791980501 (HR\,1099). If the 1st- and 2nd-order fluxed spectra differ by more than 10\%, this indicates that the observation targeting a bright source is piled up. Pileup occurs when two or more events arrive at the same (or neighboring) pixel during the same readout frame. Unfortunately, there is not much to do to alleviate the issue in existing observations \citep{Ness2007}. To mitigate the pile-up effect, proposers can request the RGS small window configuration. In this mode, only a quarter of the RGS FOV along the cross-dispersion direction will be used. Accordingly, the readout time is reduced by a factor of 4. 

\subsection{Special guide for RGS data reduction}
\label{sct:rgs_dr_special}
In the following, we provide some guidance to handle special cases: multiple X-ray bright sources in the RGS field of view (Section~\ref{sct:multi_src_fov}) and line broadening for spatially extended sources (Section~\ref{sct:ext_src}). 

\subsubsection{Multiple X-ray bright sources in the RGS field of view}
\label{sct:multi_src_fov}
\begin{figure}[!h]
\centering
\includegraphics[width=\hsize, trim={0.5cm 0.0cm 0.5cm 0.5cm}, clip]{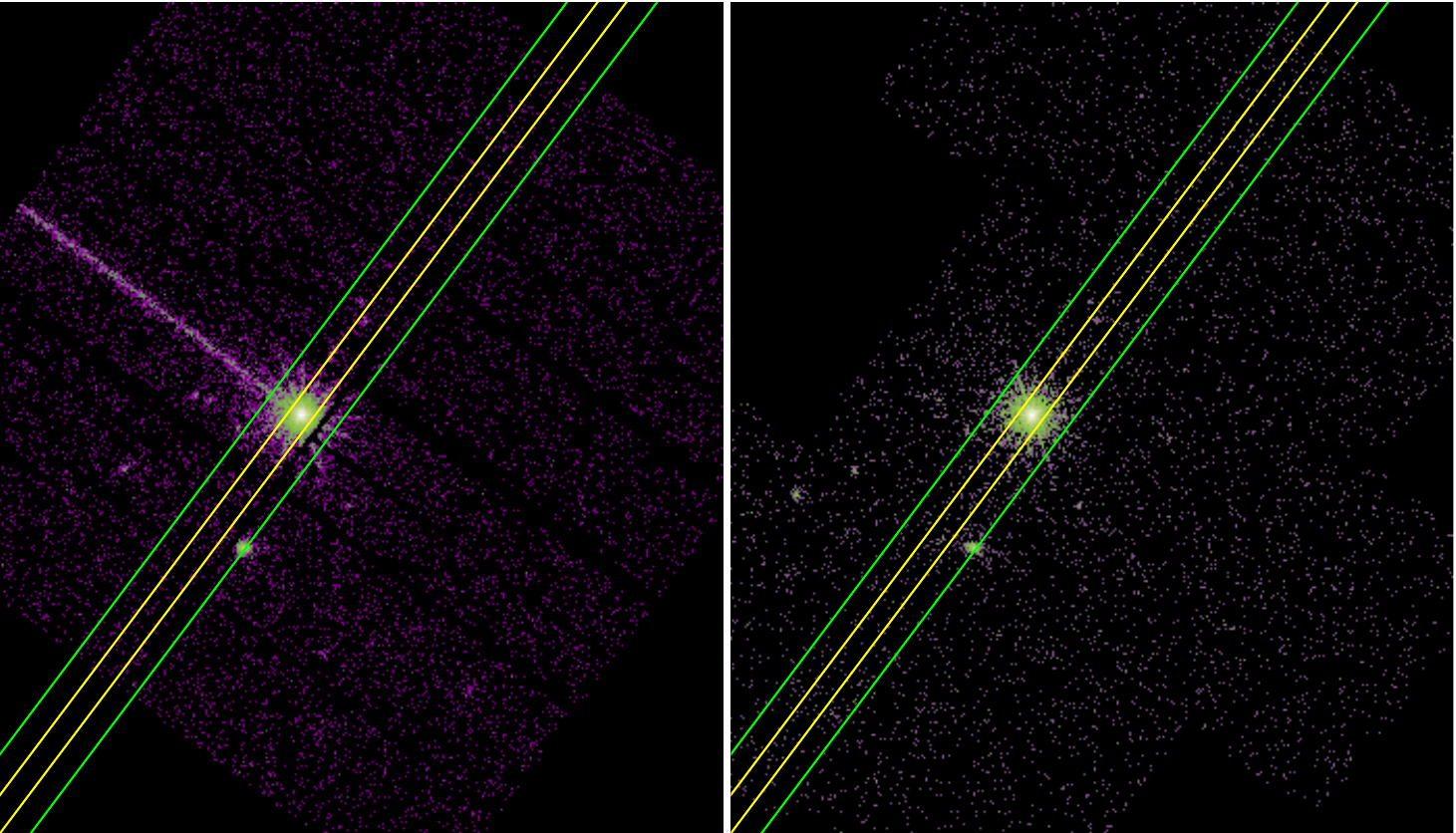}
\caption{EPIC images of Obs.ID$=$0601781401 (Mrk\,817). EPIC pn and MOS1 images are on the left and right panels, respectively. The RGS source extraction region ($\sim0.8$~arcmin along the cross-dispersion direction) is shown as tilted boxes in yellow. The X-ray bright star RX\,J1436.6+5843 also falls in the field of view of RGS (tilted green boxes). The roll angle of RGS ($\sim143^{\circ}$ in Fig.~\ref{fig:0601781401_epic_img}) is given by the fits header keyword PA\_PNT of the RGS source list file (P\$\{obsid\}R\$\{rgsid\}\$\{expid\}SRCLI\_0000.FIT).}
\label{fig:0601781401_epic_img}
\end{figure}

In some cases, there might be more than one X-ray bright source in the RGS field of view. If these sources are well-separated along the cross-dispersion direction, they might be dealt with. For instance, Obs.ID$=0601781401$, targeting Mrk\,817, has an X-ray bright star (RX\,J1436.6+5843) nearby. This star falls in the field of view of RGS (Fig.~\ref{fig:0601781401_epic_img}). To exclude this star from the background region, users need to first identify it based on its coordinates in the EPIC source list generated by the \textit{edetect\_chain} task\footnote{https://xmm-tools.cosmos.esa.int/external/sas/current/doc/edetect\_chain/edetect\_chain.html}. Subsequently, users need to run the \textit{rgsproc} pipeline with the following parameters: 

\begin{svgraybox}
\noindent user\$ \# Assuming the EPIC source list is called emmlist.fits 

\noindent user\$ \# and the source (RX\,J1436.6+5843 here) 

\noindent user\$ \# to be excluded from the background region 

\noindent user\$ \# has the index of 3 in the source list. 

\noindent user\$ rgsproc orders='1 2' withepicset=yes epicset=emllist.fits 

exclsrcsexpr='INDEX==3' 

\end{svgraybox}

\begin{figure}[h]
\centering
\includegraphics[width=.5\hsize, trim={0.5cm 0.0cm 0.5cm 0.5cm}, clip]{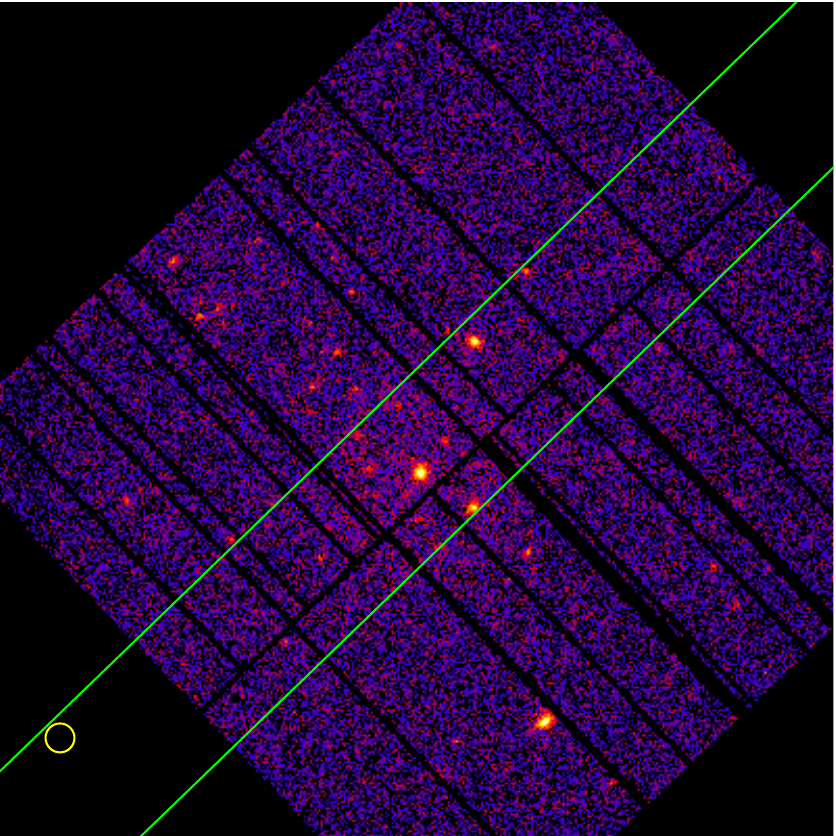}
\caption{EPIC images of Obs.ID$=$0158160201 (GRB\,031203). The RGS field of view (tilted green box) extends beyond the field of view of EPIC/pn along the dispersion direction. The yellow circle indicates the position of zeta Puppis, which is an X-ray bright star. RGS spectra of zeta Puppis can still be obtained here by setting the target coordinates properly. }
\label{fig:0158160201_epic}
\end{figure}

Caution that the RGS field of view along the dispersion direction extends beyond the field of view of EPIC. In Obs.ID$=$0158160201 (targeting GRB\,031203), the standard RGS local background spectrum (as shown in XSA) will be contaminated by the X-ray bright star zeta Puppis (Fig.~\ref{fig:0158160201_epic}). 

\subsubsection{Line broadening for extended sources}
\label{sct:ext_src}
Emission lines of HR\,1099 in Fig.~\ref{fig:plot_spec_rgs_fold} or \ref{fig:plot_spec_rgs_unfold} are narrow. The line broadening is limited by the spectral resolution of the instrument, which is $\sim0.06-0.07$~\AA\ for RGS in the first order.  

\begin{figure}
\centering
\includegraphics[width=.5\hsize, trim={0.5cm 0.0cm 0.5cm 0.5cm}, clip]{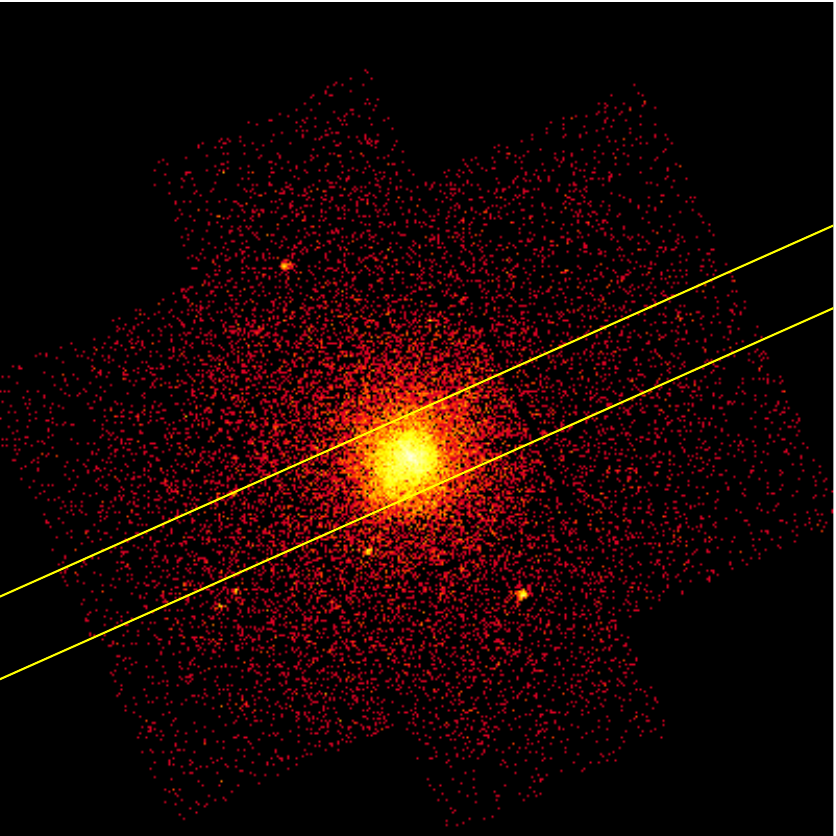}
\caption{EPIC/MOS1 images of Obs.ID$=$0037950101 (NGC\,5044). The RGS source extraction, $\sim3.4$~arcmin along the cross dispersion direction, is shown as the tilted yellow box. }
\label{fig:0037950101_epic_img}
\end{figure}

\begin{figure}
\centering
\includegraphics[width=\hsize, trim={0.5cm 0.0cm 0.5cm 0.5cm}, clip]{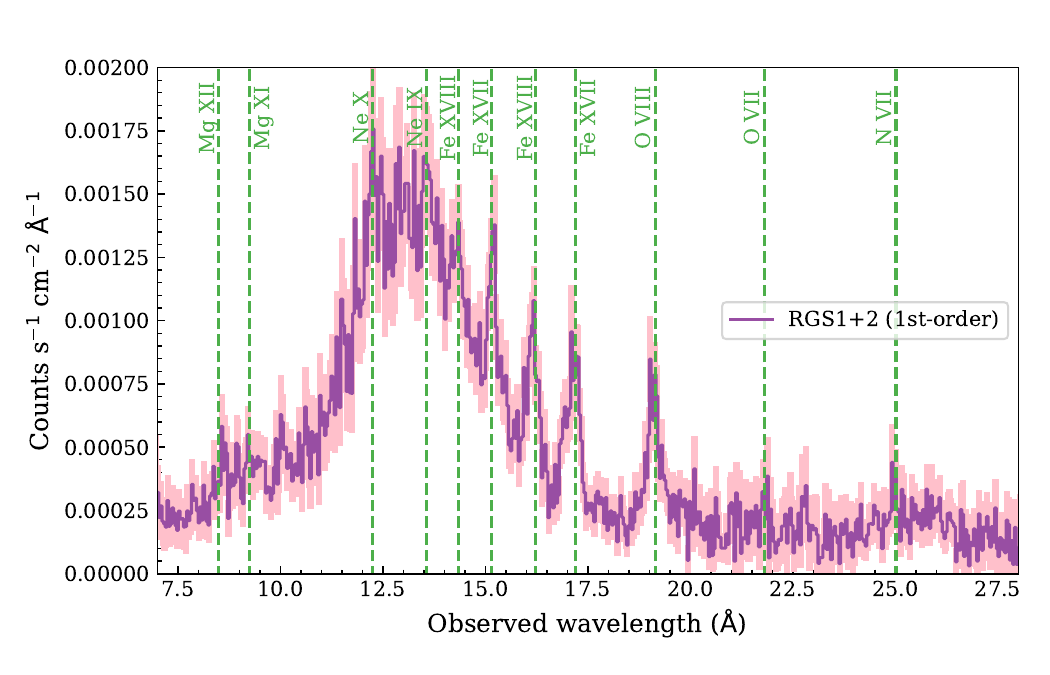}
\caption{First-order RGS spectrum (RGS1 and RGS2 combined) of Obs.ID$=$0037950101 (NGC\,5044). The source extraction region is 99\% along the cross-dispersion direction. Spectral lines are broadened due to spatial broadening.}
\label{fig:plot_spec_rgs_NGC5044}
\end{figure}

For extended sources like NGC\,5044 (see also \citep{Mao2019a}), spatial broadening dominates (Eq.~\ref{eq:dlam_rgs}). Whether the target is an extended source can be verified from the RGS or EPIC images (Fig.~\ref{fig:0791980501_rgs_img} and Fig.~\ref{fig:0037950101_epic_img}), where the source region (if symmetric) extends well beyond the 90~\% PSF ($\sim0.8$~arcmin). In this case, emission lines appear to be broader in Fig.~\ref{fig:plot_spec_rgs_NGC5044}. To be more specific, emission lines are broadened by \citep{Tamura2004}
\begin{equation}
\label{eq:dlam_rgs}
    \frac{\Delta \lambda}{\rm \AA} = \frac{0.138}{m} \frac{\Delta \theta}{\rm arcmin},~
\end{equation}
where $m$ is the spectral order, $\Delta \theta$ the spatial extent (in arcmin) of the source \citep{Tamura2004}. 


To account for such kind of spatial broadening, users can adapt the RGS response file according to the spatial extent of a moderately extended source ($\sim1$~arcmin) via the \textit{ftrgsrmfsmooth}\footnote{https://heasarc.gsfc.nasa.gov/lheasoft/ftools/fhelp/ftrgsrmfsmooth.html}. This tool is developed by Andy Rasmussen of the Columbia University XMM-Newton RGS instrument team. Alternatively, users can take advantage of the \textit{rgsvprof}\footnote{https://spex-xray.github.io/spex-help/tools/rgsvprof.html} tool. This tool is part of the SPEX code \citep{Kaastra1996}.

\section{Summary}
Since the launch of XMM-Newton in 1999, RGS has delivered thousands of high-quality high-resolution (soft) X-ray spectra \citep{Mao2019c}. Considering both the effective area and spectral resolution (Figures~\ref{fig:aeff_arcus} and \ref{fig:fom_abs_arcus}), RGS will keep playing an important role even in the era of XRISM. 

\verb|acknowledgement| 
We would like to thank Randall K. Smith for providing figures related to Arcus, Rosario Gonzalez-Riestra for valuable inputs, Guan-Fu Liu and Chunyang Jiang for feedback after careful reading. This work is based on observations obtained with XMM-Newton, an ESA science mission with instruments and contributions directly funded by ESA Member States and the USA (NASA). SRON is supported financially by NWO, the Netherlands Organization for Scientific Research.






\end{document}